\documentclass[12pt]{elsart}
\usepackage{graphicx}
\usepackage{amsmath}
\usepackage{hyperref}
\def\om{\omega}
\def\be{\begin{eqnarray}}
\def\ee{\end{eqnarray}}

\def\rmF{{\rm F}}
\def\rmd{{\rm d}}

\newcommand{\lsim}{\stackrel{\scriptstyle <}{\phantom{}_{\sim}}}
\newcommand{\gsim}{\stackrel{\scriptstyle >}{\phantom{}_{\sim}}}
\usepackage{color}

\begin{document}
	
\begin{frontmatter}
\title{Charged $\rho$-meson condensation in neutron stars
}
\author[UMB,JINR]{E.E.~Kolomeitsev},
\author[JINR,MEPHI]{K.A.~Maslov} \and
\author[JINR,MEPHI]{D.N.~Voskresensky}
\address[UMB]{Matej Bel University,  SK-97401 Banska Bystrica, Slovakia}
\address[JINR]{Joint Institute for Nuclear Research, RU-141980 Dubna, Moscow region, Russia}
\address[MEPHI]{National Research Nuclear  University (MEPhI), RU-115409 Moscow, Russia}
\begin{abstract}
We extend relativistic mean-field models with hadron masses and meson-baryon coupling constants dependent on the scalar  field $\sigma$, including hyperons and $\Delta(1232)$ baryons, to incorporate a possibility of the charged $\rho$ meson condensation in neutron star matter.  The influence of the $\rho^-$ condensation on the equation of state proves to be strongly model dependent. In our models of one type (KVORcut-based ones) the $\rho^-$ condensation arises by a second-order phase transition above a critical density and   the maximum value of the  neutron star mass diminishes only slightly. The matter composition changes more significantly. In our models of other type (MKVOR*-based ones),  if the system is considered at  fixed density, the $\rho^-$ condensation  arises by a second-order phase transition  at the baryon density $n=n_{c,\rho}^{(\rm II)}$ and at a slightly higher density $n=n_{c,\rho}^{(\rm I)}$ there  occurs a first-order phase transition. In a neutron star matter starting with  a  density $n<n_{c,\rho}^{(\rm II)}$ there appears a region of a  mixed phase, or the system is described by  Maxwell construction, that   results in a substantial decrease of  the value of the maximum neutron star mass. Nevertheless in the models under consideration  the observational constraint on the maximum neutron star mass is  fulfilled. Besides, in MKVOR*-based models the appearance of the $\rho^-$ condensate  is accompanied by a strong rearrangement of the matter composition. Dependence of the results on a choice of the $\rho$ meson scaling functions for the effective $\rho$ meson mass and coupling constants is also investigated.
\end{abstract}
\end{frontmatter}


\section{Introduction}

A nuclear equation of state (EoS) is one of the key ingredients in the description of neutron star (NS) properties~\cite{Lattimer:2012nd}, supernova explosions~\cite{Woosley} and heavy-ion collisions~\cite{Danielewicz:2002pu,Fuchs}.
A rather popular class of hadronic models is  the relativistic mean-field (RMF) approach~\cite{Walecka1974,Boguta77,Boguta77a,Boguta77b,Boguta77c,SerotWalecka,SerotWaleckaA,Glendenning,Weber,Savushkin2015}.
As any phenomenological model, RMF models have to be gauged by some empirical information or more founded microscopic calculations.

Any reliable EoS of the cold hadron matter should satisfy experimental information on global properties of dilute nuclear matter and atomic nuclei, agree with the results of more microscopic approaches~\cite{APR,FP,Gandolfi:2009nq,Gandolfi12,Gandolfi12a,Hebeler:2014ema,Tews13} for densities $n\lsim 1.5\mbox{--}2 \,n_0$ ($n_0 = 0.16$ fm$^{-3}$ is the saturation nuclear density), not contradict constraints on the pressure of the nuclear mater  extracted from the description of particle transverse and elliptic flows~\cite{Danielewicz:2002pu} and the $K^+$ production~\cite{Lynch} in heavy-ion collisions, allow for the heaviest known compact stars PSR~J1614-2230 with the mass $1.97\pm 0.04 M_{\odot}$ (this result of~\cite{Demorest:2010bx} was recently corrected in~\cite{Fonseca} and now reads as $1.928\pm 0.017 M_{\odot}$) and PSR~J0348+0432 with the mass $2.01\pm 0.04 M_{\odot}$~\cite{Antoniadis:2013pzd} ($M_\odot$ is the solar mass), allow for an adequate description of the NS cooling, cf.~\cite{Blaschke:2004vq,Kolomeitsev:2004ff,Grigorian:2016leu},
yield a mass-radius relation comparable with the empirical constraints~\cite{Bogdanov:2012md,Hambaryan2014,Heinke:2014xaa}. Besides, there are some other constraints suggested in the literature, which perhaps should be also fulfilled, e.g. the constraint on the gravitational mass and total baryon number of the pulsar PSR~J0737-3039(B)~\cite{Podsiadlowski,Kitaura:2005bt}, see also discussion in~\cite{Blaschke2016}, and the unitary gas constraint on the EoS~\cite{Kolomeitsev2016}.
The constraints on the high density part of the nuclear EoS were discussed in~\cite{Klahn:2006ir}. Some of the  constraints considered in~\cite{Klahn:2006ir} were recently tightened and new constraints were added.
Additionally, the so-constrained EoS when extended to non-zero temperatures should describe  heavy-ion collisions~\cite{Khvorostukhin:2006ih,Khvorostukhin:2008xn}.

It turns out to be the most difficult to reconcile the constraint on the maximum compact star mass, $M_{\rm max} > 1.97\,M_\odot$, cf.~\cite{Demorest:2010bx,Antoniadis:2013pzd}, and the constraints on the stiffness of the EoS extracted from the analyses of flows in heavy-ion collisions~\cite{Danielewicz:2002pu,Fuchs}. This dichotomy sharpens if the appearance of other baryonic species besides nucleons is allowed in a model. With inclusion of any new species the EoS becomes softer and the maximum NS mass reduces. Information about hyperon-nucleon interactions in vacuum and properties of hypernuclei implies a possibility for hyperons to appear in a NS. Accordingly, in standard models of EoS, like RMF non-linear Walecka models, there arises a strong reduction of the maximum NS mass, see~\cite{Glendenning,Djapo:2008au,Schulze-Rijken}. Similar problem  appears, if $\Delta(1232)$ baryons are incorporated~\cite{Drago2014,Drago:2015cea}. In this context one speaks about the ``hyperon puzzle'' and the ``$\Delta$ puzzle''.

To make RMF models more flexible and to reconcile nucleon self-energies obtained in the mean-field approximation with the results of more microscopic Dirac-Bruckner-Hartree-Fock calculations based on the realistic nucleon-nucleon potentials
a family of RMF models with the baryon density dependent meson-nucleon coupling constants was developed, cf.~\cite{Long,Fuchs,Typel,Hofmann,Niksirc,Lalazisis,Gaitanos,RocaMaza:2011qe}.
A detailed analysis of such models  is performed in~\cite{Typel2005,Typel2005a}.

Reference~\cite{Kolomeitsev:2004ff} introduces  RMF models incorporating in-medium modifications of the baryon and meson effective masses and coupling constants considering them as $\sigma$-field dependent. This was motivated by experimental hints on modification of hadronic properties in hadronic matter like changes of the hadron masses and widths, cf.~\cite{Metag}, and by theoretical arguments that  masses of all hadrons except Goldstone bosons, like pions and kaons, should decrease with increasing density and/or temperature due to a partial restoration of the chiral symmetry in dense and/or hot matter, cf.~\cite{Rapp,Koch}. Brown and Rho conjectured in~\cite{BrownRho,BrownRhoA} that the nucleon effective mass and the effective masses of vector $\omega$, $\rho$ and scalar $\sigma$ mesons should scale approximately universally in dense nucleon matter. In the framework of the RMF approach for infinite matter the effective hadron masses ($m^{*}_h$) and the coupling constants ($g^{*}_h$) enter all relations only in combinations $m^{*\,2}_h/g^{*\,2}_h$ that leads to equivalence theorem between  different  RMF schemes~\cite{Kolomeitsev:2004ff}. Allowing for differences among scaling functions for the effective hadron masses and coupling constants one can better fulfill various experimental constraints on the EoS.
One of the constructed  models labeled then in~\cite{Klahn:2006ir} as the KVOR model fulfilled appropriately most of the constraints known to that time. However the maximum NS mass in the KVOR model only slightly exceeds the recently measured mass of the pulsars PSR~J1614-2230  and PSR~J0348+0432. Hyperons and $\Delta$-isobars were not incorporated.

In~\cite{Maslov:2015msa,Maslov:2015wba} we proposed two modifications of the KVOR model.
We use the stiffening method, described in~\cite{Maslov:cut}, applicable to an arbitrary EoS in RMF models. It was observed that the EoS stiffens, if at baryon densities exceeding a certain value above $n_0$ growth of the scalar-field magnitude with the density is bounded from above at some value. In the KVORcut modifications of the KVOR model, it was realized by making a sharp change of the nucleon coupling constant to the vector meson as a function of the scalar field slightly above the desired value. In another version of the model, labeled as MKVOR model,  quenching of the scalar field is implemented in the isospin-asymmetric matter with the help of a strong variation of the effective nucleon coupling constant  to the isovector meson as a function of the scalar field. Therefore, the equation of state becomes stiffer in asymmetric NS matter, while staying soft in isospin-symmetric matter (ISM). Also it assumes a smaller value of the nucleon effective mass at the nuclear saturation density. Both these  models incorporated hyperons. Most of the constraints on EoS, including  maximum NS mass constraint, were appropriately fulfilled.

In the recent work \cite{MKVOR-Delta} we additionally included $\Delta$ resonances in the KVORcut03 and MKVOR-based models with $\sigma$-scaled hadron masses. It proved to be that in the MKVOR$\Delta$ model the effective nucleon mass in ISM drops to zero at $n\sim (4-6)n_0$, if one exploits a relevant value for the $\Delta$  potential $U_\Delta (n_0) \sim -(50-100)$~MeV. Then within this model the hadronic EoS cannot be used at a higher density and should be replaced by the quark one. In order to continue dealing with the hadronic description we slightly modified the MKVOR model, labeled  as MKVOR*, that exploits the cut-mechanism both in $\rho$ and  $\omega$ sectors. Then the effective nucleon mass does not reach zero. We demonstrated that within the KVORcut03- and MKVOR*-based models one is still able  to construct an appropriate EoS with inclusion of hyperons and $\Delta$s, satisfying presently known experimental constraints including  the NS maximum mass constraint.

Besides the appearance of the hyperon and $\Delta$ admixtures in the baryon Fermi sea, various phase transitions to the states of Bose condensates, like pion, kaon  and charged $\rho$ meson condensates, may appear in dense NS interiors, resulting in an additional softening of the EoS and causing extra troubles with the fulfillment of the maximum NS mass constraint. All these phase transitions result in similar consequences. In this paper we will focus on  the charged $\rho$ condensation.
The latter possibility was suggested in~\cite{v97}  and then studied in~\cite{Kolomeitsev:2004ff}.  Within our RMF models, where the  $\rho$ meson is included  as a non-Abelian gauge boson, the charged $\rho$ condensation may occur owing to decrease
of the effective $\rho$ meson mass  with increase of the baryon density.

Our work is organized as follows. In Section~\ref{sec:model} we formulate the energy density functional for our generalized RMF model with scaled hadron masses and couplings with inclusion of hyperons,  $\Delta$ isobars and charged $\rho$ condensation. In Section~\ref{EoSdeltas} we investigate KVORcut03 and MKVOR*-based  models with inclusion of hyperons, $\Delta$ baryons and $\rho^-$ condensate (KVORcut03H$\Delta\phi\rho$ and MKVOR*H$\Delta\phi\rho$, respectively). In Section \ref{variat} we investigate effects of variations of scaling functions in MKVOR*-based models.
The results are summarized in the Conclusion. Technical details are collected in Appendices~A--C.

\section{Energy density functional}\label{sec:model}

\subsection{General formulations}

The Lagrangian of the model we employ is formulated in~\cite{Maslov:2015wba,MKVOR-Delta}, see Appendix~\ref{app:Lag}  for a short review. The model is a generalization of the non-linear Walecka model with effective coupling constants $g_{mb}^{*} = g_{mb} \chi_{mb}(\sigma)$ and hadron masses $m_i^{*} = m_i \Phi_i(\sigma)$, which are functions of the $\sigma$ field~\cite{Kolomeitsev:2004ff}, $i=\{b,m\}$, where $m = \{\sigma, \omega, \rho, \phi\}$ lists the included mesonic fields,  $b=\{N,H,\Delta\}$  indicates a baryon (nucleon $N=p,n$; hyperon $H=\Lambda,\Sigma,\Xi$; and $\Delta$ isobar). Quantities $\chi_{mb}(\sigma)$ and  $\Phi_i(\sigma)$ are dimensionless scaling functions of the $\sigma$ mean field. In the RMF approximation the contribution of $\Delta$s to the energy density has the same form as for spin-$1/2$ fermions but with the spin degeneracy factor 4.
With the standard solutions for the $\om$ and $\rho$ meson mean fields, being expressed through baryon densities, the energy density  of our model takes the following form
\begin{align}
\label{edensity}
E[{\{n_b\}},\{n_l\},  f] &=  \sum_b E_{\rm kin}(p_{{\rm F},b}, m_b\,\Phi_b(f), s_b)
+
\sum_{l=e,\mu} E_{\rm kin}(p_{{\rm F},l},m_l, s_l)
\nonumber \\
&
+\frac{m_N^4 f^2}{2 C_\sigma^2 }\eta_\sigma(f)
+\frac{1}{2 m_N^2} \Big[\frac{C_\om^2 n_V^2}{ \eta_\om(f)}
+\frac{C_\rho^2 n_I^2}{\eta_\rho(f)}
+ \frac{C_\phi^2 n_S^2}{\eta_\phi(f)}
\Big] , \,
\\
n_V =\sum_b x_{\om b} n_b\,,&\quad
n_I =\sum_b x_{\rho b} t_{3 b} n_b\,,\quad
n_S =\sum_{H} x_{\phi H} n_H\,,
\nonumber
\end{align}
where we introduced the dimensionless scalar field $f = {g_{\sigma N}\chi_{\sigma N}(\sigma) \sigma}/{m_N}$, the coupling constant ratios $x_{Mb}=g_{Mb}/g_{MN}$, $M = \{\sigma, \om, \rho\}$, $x_{\phi H}=g_{\phi H}/g_{\om N}$, $g_{\phi N}=g_{\phi \Delta}=0$, and $t_{3b}$ is the isospin projection of baryon $b$; the Fermi momentum is related to the fermion density  $p_{{{\rm F},j}} = (6 \pi^2 n_j / (2 s_j + 1))^{1/3}$ with $s_j$ standing for spin of fermion $j=(b,l)$, $l=(e,\mu)$;
\begin{align}
C_M = \frac{g_{MN} m_N}{m_M},  \quad
C_\phi = \frac{g_{\om N}\, m_N}{m_\phi}\,.
\end{align}
The scaling functions are
\begin{gather*}
\eta_\om (f) = \frac{\Phi^2_\om (f)}{\chi^2_{\om N}(f)}, \, \eta_\rho (f) = \frac{\Phi^2_\rho (f)}{\chi^2_{\rho N}(f)}, \, \eta_\phi (f) = \frac{\Phi^2_\phi (f)}{\chi^2_{\phi H}(f)}, \\
\eta_{\sigma}(f)=\frac{\Phi_{\sigma}^2[\sigma(f)]}{\chi_{\sigma N}^2[\sigma(f)]} + \frac{ 2 \, C_{\sigma}^2}{m_N^4 f^2}  {U}[\sigma(f)]\,.
\end{gather*}

As in~\cite{Maslov:2015msa, Maslov:2015wba, MKVOR-Delta}, we use universal mass scaling for nucleons and mesons, $\Phi_N=\Phi_m =1-f$, and universal scaling for coupling constants
$\chi_{\omega b}(f) = \chi_{\omega N}(f)$ and $\chi_{\rho b}(f)
= \chi_{\rho N}(f)$, the coupling constant for $\phi$  meson remains unscaled, $\chi_{\phi H}(f) =1$, that results in the $\phi$ scaling function $\eta_\phi=(1-f)^2$.
 The scaling $\eta_{\sigma}(f)$ includes also the scalar meson self-interaction potential ${U}(\sigma)$ entering the Lagrangian of the model (\ref{Lag-mes}).
The coupling constant of the scalar field to hyperons and $\Delta$s may differ from that of the scalar field to nucleons. Taking this into account the scaling function of the baryon mass can be written as
\begin{align}
\Phi_b(f) = 1-g_{\sigma b}\chi_{\sigma b}\sigma =1 - x_{\sigma b}\frac{m_N}{m_b} f,
\nonumber
\end{align}
where we assume the universality $\chi_{\sigma b}=\chi_{\sigma N}$.
Expressions for the scaling functions $\eta_{m}(f)$ for the KVORcut03 and MKVOR* models adjusted in~\cite{Maslov:2015msa, Maslov:2015wba, MKVOR-Delta} to achieve the best description of various constraints on the nuclear EoS are collected in Appendix~\ref{app:scale-fun}.
The fermion kinetic energy density is
\begin{align}
E_{\rm kin}(p_{\rm F},m,s) = (2s + 1)\intop_0^{p_{{\rm F}}}
\frac{p^2 \rmd p}{2 \pi^2} \sqrt{p^2 + m^2}.
\end{align}

Under the ISM, which exists on short time scales when weak processes are not operative, we  understand the hyperon-free matter with $n_I=0$. Then the  chemical potentials of $\Delta$ baryons and nucleons are related as
\begin{align}
\mu_{\Delta^{++}}=\mu_{\Delta^+}=\mu_{\Delta^0}=\mu_{\Delta^-}=\mu_n=\mu_p\,.
\end{align}
Composition of the NS matter is determined by weak processes bringing the matter in beta-equilibrium. The particle densities as functions of the total baryon density, $n=\sum_b n_b$, in beta-equilibrium matter (BEM) follow from the conditions
\begin{align}
\mu_b = \mu_n - Q_b \mu_e,
\label{chempot-i}
\end{align}
where  $Q_b$ is the electric charge of the baryon,
$\mu_j = \frac{\partial  E}{\partial n_j}$  are the baryon and the lepton ($e,\mu$) chemical potentials, the condition $\mu_e=\mu_\mu$, and the charge neutrality condition
\begin{align}
\sum_b Q_b n_b - n_e - n_\mu = 0\,.\label{electroneut}
\end{align}
These equations are solved together with the equation of motion for the scalar field
$\partial E/ \partial f = 0$.

Finally the pressure as a function of the baryon density and corresponding concentrations can be calculated as
\begin{align}
P = \sum_{b} \mu_b n_b + \sum_{l} \mu_l n_l - E[{\{n_b\}},\{n_l\}, f(n)]\,.
\label{press}
\end{align}

The hyperon and $\Delta$ couplings with vector mesons are related to the nucleon ones via the SU(6) symmetry relations:
\begin{align}
x_{\om \Lambda} =x_{\om \Sigma}= 2x_{\om \Xi} = {\textstyle\frac{2}{3} } , \quad
x_{\rho \Sigma} = 2 x_{\rho \Xi} = 2  \,,\quad g_{\rho\Lambda}=g_{\phi N} = 0\,,
\nonumber\\
2 g_{\phi \Lambda} = 2 g_{\phi \Sigma} = g_{\phi \Xi} = - {\textstyle\frac{2\sqrt{2}}{3} }g_{\om N} ,
\quad
x_{\omega \Delta} = 1, \quad x_{\rho \Delta} = 1, \quad g_{\phi \Delta} = 0\,.
\nonumber
\end{align}
Coupling constants of baryons with the $\sigma$ meson are found from the baryon potentials in ISM at the saturation density,
\begin{align}
U_{b} = \frac{C_{\omega}^2}{\eta_\om(f_0)}\frac{n_0}{m_N^{2}} x_{\omega b}  - (m_N-m_N^{*} (n_0))x_{\sigma b},
\end{align}
where $f_0$ is the solution of equation of motion in the ISM at saturation, $n_p = n_n = n_0/2$, and we set $U_{\Lambda } = -28 \,{\rm MeV}$, $U_{\Sigma } = 30 \,{\rm MeV}$, $U_{\Xi } = -15 \,{\rm MeV}$. As argued in~\cite{MKVOR-Delta}
the realistic value of the $\Delta$ potential is close to the nucleon one $U_N \sim -(50-60) {\rm MeV}$ and in this paper we will use the value $U_\Delta\simeq -50$\,MeV that was found in~\cite{Riek:2008uw} from the analysis of the pion photoproduction off nuclei.

The parameters of the models, which we exploit in the present work are given in Appendix~\ref{app:scale-fun}.

\subsection{Phase transition to charged $\rho$ condensate}\label{sec:rho}

In the framework of the hidden local symmetry model~\cite{Bando,Bando1} the self-interacting $\rho$ meson field is introduced as a non-Abelian field.
The self-interaction of the non-Abelian gauge fields may lead to  appearance of charged condensates in strong fields in QCD~\cite{Migdal78qcd}  (gluon condensate) and in the electro-weak sector of the Standard Model~\cite{Linde79}  ($W$ boson condensate).
The non-Abelian $\rho$ meson field is analogous to the massive gluon field. Reference~\cite{v97} demonstrated that in dense nucleon isospin-asymmetric matter at an appropriate decrease of the effective $\rho$ meson mass there occurs a charged $\rho$ condensation. Then the phenomenon of the charged $\rho^-$ condensation was incorporated in the RMF models with scaled hadron masses and coupling constants in~\cite{Kolomeitsev:2004ff}. Similarly, the charged $\rho^-$ condensation may appear in very strong magnetic fields~\cite{Chernodub,Mallick:2014faa}.

Considering a possibility of the non-vanishing charged $\rho$ fields we have to rewrite the $\rho$ meson sector of the Lagrangian, see (\ref{Lag-mes}) in Appendix~\ref{app:Lag}, taking into account the chemical potential, $\mu_{{\rm ch},\rho}$, for charged mesons,
\begin{eqnarray}
\mathcal{L}_\rho &=& -
\frac14 \vec{R}_{\mu \nu}\vec{R}^{\mu \nu}
+ \frac12 m_\rho^{2}\Phi_\rho^2 \vec{\rho}_\mu \vec{\rho}^{\,\mu} - \sum_b g_{\rho b} \chi_{\rho b} \bar{\Psi}_b\gamma^\mu \vec{t}_b \vec{\rho}_\mu \Psi_b\,,
\label{Lag-rho}\\
\vec{R}_{\mu \nu} &=&
\partial_\mu \vec{\rho}_\nu -  \partial_\nu \vec{\rho}_\mu + g_\rho' \chi_\rho' [\vec \rho_\mu \times \vec \rho_\nu]
+ \mu_{{\rm ch},\rho} \delta_{\nu 0} [\vec n_3 \times \vec \rho_\mu] - \mu_{{\rm ch},\rho} \delta_{\mu 0} [\vec n_3 \times \vec \rho_\nu]
\,,
\nonumber
\end{eqnarray}
where $(\vec{n}_3)^a=\delta^{a3}$ is the unit vector in the isospin space;  $\chi_{\rho b}$, $\chi_\rho'$ are the corresponding scaling functions of coupling constants.

The  ansatz for the $\rho$ meson fields~\cite{v97} includes $\rho_0^{(3)}$ field and spatial components of  charged $\rho$ meson fields, $\rho_i^{(\pm)} =(\rho_i^{(1)} \pm i\rho_i^{(2)})/\sqrt{2}\neq 0$\,, $i=1,2,3$\,. In the absence of strong  magnetic fields the fields $\rho_i^{(3)}$, $\rho_0^{(i)}$ lead to an increase of the energy of the system and should, therefore, be put zero, and  the condition $$\rho_i^{(+)}\rho_j^{(-)}-\rho_i^{(-)}\rho_j^{(+)}=0$$
minimizes the energy. This implies that the ratio $\rho_i^{(+)}/\rho_i^{(-)}$ is constant independently  of the spatial index $i$. Then we may take $\rho_i^{(-)} = a_i\,\rho_c$ and $\rho_i^{(+)}=a_i\, \rho_c^\dagger$, where $\vec{a}=\{a_i\}$ is a spatial unit vector, and $\rho_c$ is a complex amplitude of the charged $\rho$ meson field. Note that this state corresponds to the zero  average spin of the $\rho$ meson field.
With the above ansatze we have
\begin{eqnarray}
&R^{(\pm)}_{\mu\nu}=\frac{1}{\sqrt{2}}(R^{(1)}_{\mu\nu}\pm i R^{(2)}_{\mu\nu})
=\pm i (g_\rho^{'}\chi'_\rho \rho_0^{(3)}-\mu_{{\rm ch},\rho})
[\rho_\mu^{(\mp)}\,\delta_{\nu 0}-\delta_{\mu 0}\,\rho_\mu^{(\mp)}]
\nonumber
\end{eqnarray}
and $R^{(3)}_{\mu\nu}=0\,.$ Therefore,
$$-\quart \vec{R}_{\mu\nu}\vec{R}^{\mu\nu}=-\half R_{\mu\nu}^{(+)} R^{(-),\mu\nu}= (g_\rho^{'}\chi'_\rho \rho_0^{(3)}-\mu_{{\rm ch},\rho})^2\,|\rho_c|^2\,.$$

For the $\rho$ meson included according to the hidden local symmetry principles~\cite{Bando,Bando1} one has $g_{\rho N} =g_{\rho}^{'} \equiv g_\rho$. Despite this, there  may be $\chi_{\rho N}\neq \chi_{\rho}^{'}$  in dense nuclear matter. Indeed, the scaling factor $\chi_{\rho N}$ arises due to the renormalization of the interaction between nucleon and $\rho$ meson  in medium, whereas the factor $\chi_\rho^{'}$ is due to a non-Abelian
interaction between $\rho^\alpha_i$ fields and does not depend directly on the nucleon source. We will continue to use $\chi_{\rho N}=\chi_{\rho b}$. Then
the contribution to the thermodynamic potential density ($\Omega =-P$) in the $\rho$ sector can be written as a function of $\rho_0^{(3)}$ and $\rho_c$ fields in the following form~\cite{Kolomeitsev:2004ff}:
\begin{align}
\Omega_\rho[\{n_b\};f,\rho_0^{(3)}, \rho_c] &=
g_{\rho}\, {\chi}_{\rho N} n_I\, \rho_0^{(3)}
- \frac12 (\rho_0^{(3)} )^2\, m_\rho^{2}\, \Phi_\rho^2
\nonumber\\
&
-\left[\big(g_\rho{\chi'}_\rho \, \rho_0^{(3)}  -\mu_{{\rm ch},\rho}\big)^2
-m_\rho^{2}\, \Phi_\rho^2\right]\, |\rho_c|^2
\,.
\label{er}
\end{align}

Variation  with respect to the fields
$\rho_0^{(3)}$ and $\rho_i^-$ yields the equations of motion
\begin{align}
&\big[ \big( g_{\rho}\,{\chi}^{'}_\rho\, \rho_0^{(3)}
- \mu_{{\rm ch},\rho} \big)^2-m_\rho^2\,\Phi_\rho^2 \big]
\, \rho_c =0\,,
\label{rhoem}\\
&
m_\rho^{2}\,\Phi_\rho^2\, \rho_0^{(3)} +
2\, g_\rho\,{\chi}^{'}_\rho\,
\big(g_\rho{\chi'}_\rho \, \rho_0^{(3)} -\mu_{{\rm ch},\rho}\big)
|\rho_c|^2 \,  =  g_{\rho}\, \chi_{\rho N}\,n_I\,.
\nonumber
\end{align}
This system of equations has two solutions. The first one is the  traditional solution, $\rho_c=0$ and
$\rho_0^{(3)}= \frac{g_\rho\chi_{\rho N}}{m_\rho^2\Phi_\rho^2} n_I$,
yielding
\begin{align}
\Omega_{\rho}^{(1)}[n_n,n_p;f] =\frac{C_\rho^2 n_I^2}{2\,m^2_N\,
  \eta_\rho(f)}\,.
  \label{ERho}
\end{align}
The second possible solution of (\ref{rhoem}) is
\begin{align}
&\rho_0^{(3)} = \frac{\mu_{{\rm ch},\rho} +\zeta m_\rho\Phi_\rho }{g_\rho{{\chi}'_\rho}}
\,,\quad
|\rho_c|^2=\frac{\zeta\, n_I - n_\rho}{2\,m_\rho\, \eta^{1/2}_\rho\,{\chi}'_\rho}\,,
\nonumber\\
& n_\rho=a\,(m_\rho\,\Phi_\rho +\zeta\mu_{{\rm ch},\rho})\,, \,\, a=\frac{ m_N^2\eta^{1/2}_\rho\, \Phi_\rho}{C_\rho^2{\chi}'_\rho} >0\,,
\label{solc}
\end{align}
which exists only if $\zeta\, n_I - n_\rho>0$. The corresponding contribution to the thermodynamic potential is
\begin{align}
&\Omega_{\rho}^{(2)}[n_n,n_p;f] = \Omega_{\rho}^{(1)}[n_n,n_p;f]
 -
\frac{C_\rho^2}{2\, m_N^2\, \eta_\rho} \big(\zeta n_I -  n_\rho
\big)^2\theta(\zeta n_I -  n_\rho).
\label{enrhoc}
\end{align}
In Eqs.~(\ref{solc}) and (\ref{enrhoc}) $\zeta=\pm 1$ represents two possible solutions of the first equation in (\ref{rhoem}).
These solutions correspond to two different charges of the $\rho$ condensate as
\begin{align}
n_{{\rm ch},\rho}= \frac{\partial  \Omega_\rho }{\partial \mu_{{\rm ch},\rho}} &=
2\,\big(g_\rho{\chi'}_\rho \, \rho_0^{(3)}-\mu_{{\rm ch},\rho}\big)
|\rho_c|^2
=2\zeta m_\rho \Phi_\rho |\rho_c|^2\,.
\label{nchrho}
\end{align}
If the NS composition is such that $n_I<0$ (as usually is the case for the neutron-rich matter) then from the condition $|\rho_c|^2 >0$ follows that we must take $\zeta = -1$. Then we deal with the $\rho^-$ condensate. The beta-equilibrium is sustained by the reaction $n + e^- \leftrightarrow n + \rho^-$, which yields the equality $\mu_{{\rm ch},\rho} =\mu_e$. If $n_I$ were $>0$, we would take $\zeta = +1$ and deal with the $\rho^+$ condensate, then the reaction $p  \leftrightarrow n + \rho^+$ would establish the relation $\mu_{{\rm ch},\rho} =-\mu_e$. However such a possibility is not realized for the models, which we consider. Thereby below we deal with the $\rho^-$ condensate.

The  contribution to the energy density (\ref{edensity}) from the charged $\rho^-$ meson condensate is then given by
\begin{align}\label{de}
\Delta E_{{\rm ch},\rho}[\{n_b\};f]=- \frac{C_\rho^2}{2\, m_N^2\, \eta_\rho} \big(n_I +  n_\rho\big)^2\theta (-n_I - n_\rho)-\mu_{\rm ch,\rho}n_{\rm ch, \rho}\,,
\end{align}
$\theta (-n_I - n_\rho)=1$	for $n_I + n_\rho < 0$ and zero otherwise.

Note that following Eqs.~(\ref{enrhoc}), (\ref{de}) the charged $\rho^-$ condensate appears in a second-order phase transition at a monotonous increase of the baryon and isospin densities, cf.~\cite{Kolomeitsev:2004ff}. However a transition
to a new solution could be of a first order with an abrupt change of the isospin composition, cf.~\cite{v97}. To check the latter possibility one needs to compute the total thermodynamic potential  of the system.

We see that the appearance of the charged $\rho^-$ condensate is favored by an increase of the
magnitude of the isospin density, $|n_I|$, by a decrease of the effective $\rho$ meson mass in the medium, $m_\rho\Phi_\rho$, and by an increase of the electron chemical potential.

In the presence of the charged $\rho^-$ condensate the electric neutrality condition reads
\begin{align}
\sum_b Q_b n_b - n_e - n_\mu + n_{\rm ch,\rho} = 0.
\label{elneutcond-rho}
\end{align}
Finally the pressure is given by
\begin{align}
P &= \sum_{b} \mu_b n_b + \sum_{l} \mu_l n_l +\mu_{\rm ch,\rho} n_{\rm ch,\rho}
\nonumber\\
&-E[{\{n_b\}},\{n_l\},  f(n)] -\Delta E_{{\rm ch},\rho}[\{n_b\};f(n)]\,,
\end{align}
where the particle concentrations and the magnitude of the scalar field $f(n)$ follow from the solution of equation $\partial E/ \partial f = 0$, the electro-neutrality equation (\ref{elneutcond-rho}) and the beta-equilibrium conditions (\ref{chempot-i}), where in the latter  the chemical potential of the baryon $b$ is taken in the presence of the charged $\rho^-$ condensate,
\begin{align}
\mu_b&=\sqrt{p_{\rmF,b}^2+m_b^2\Phi_b^2({f(n)})}
+\frac{1}{m_N^2} \Big[\frac{x_{\om b}C_\om^2 n_B}{ \eta_\om({f(n)})}
-\frac{x_{\rho b}t_{3b}\,C_\rho^2 n_\rho}{\eta_\rho({f(n)})}
+ \frac{x_{\phi b} C_\phi^2 n_S}{\eta_\phi({f(n)})}
\Big].
\label{mub-rho}
\end{align}
In the absence of the charged $\rho^-$ condensate we have to replace $n_\rho \to -n_I$ in this expression.

\section{Charged $\rho^-$ condensate in KVORcut03- and MKVOR*-based models}\label{EoSdeltas}

Further on, various suffixes in the model labels denote the included baryons (H for hyperons and $\Delta$ for $\Delta$s) and the particular choice of scaling functions in the strangeness sector ($\phi$), cf.~\cite{MKVOR-Delta}. The suffix ($\rho$) indicates the models in which the $\rho^-$ condensation is included. Throughout the rest of the paper we for simplicity
assume $\chi'_\rho = 1$, as we see no any solid reason for its scaling, cf.~\cite{Kolomeitsev:2004ff}.

\subsection{KVORcut03-based models}

\begin{figure*}
\centering
\includegraphics[width=14cm]{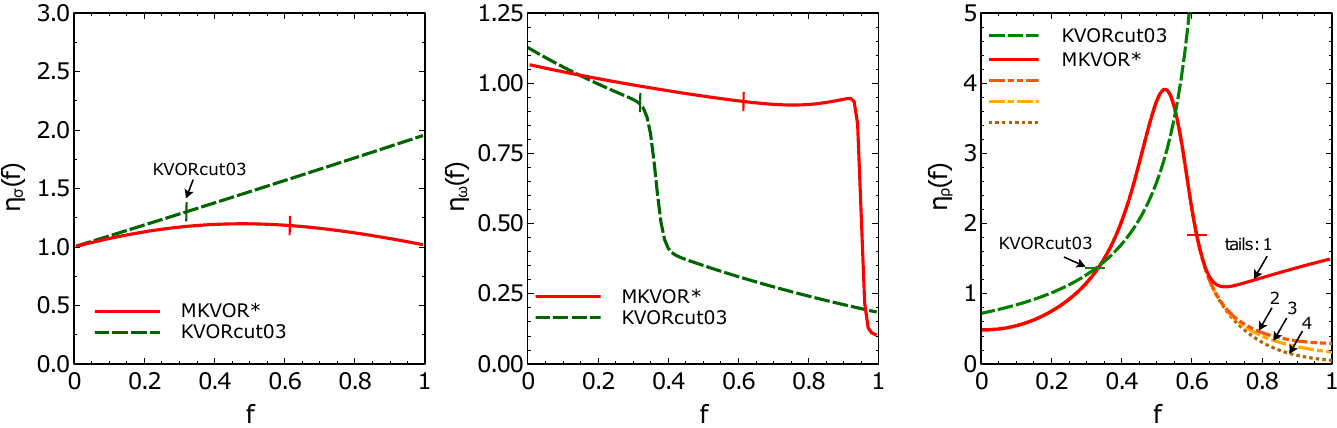}
\caption{Scaling functions $\eta_\sigma$, $\eta_\om$, $\eta_\rho$ in BEM as functions of the scalar field $f$ for KVORcut03 and  MKVOR* models. Vertical and horizontal bars indicate the maximum values of $f$ ($f_{\rm lim}$) reachable in the NS with the maximum mass for the corresponding model.}
\label{Fig-1-new}
\end{figure*}

For the  model ~\cite{Kolomeitsev:2004ff}, labeled in \cite{Klahn:2006ir} as KVOR model, the cut mechanism of ~\cite{Maslov:cut}  was applied in~\cite{Maslov:2015wba,MKVOR-Delta}.
The KVORcut03 version of the model was proved to fulfill both the flow and the maximum NS mass constraints as well as most of other existing constraints. Inclusion of the hyperons and $\Delta$s in the model did not spoil fulfilment of those constraints. Therefore here we continue to test the KVORcut03-based models including now the possibility of $\rho^-$ condensation.  Expressions for the scaling functions and values of parameters are given in Appendix~\ref{app:scale-fun}. Scaling functions $\eta_\sigma (f)$, $\eta_\om (f)$, $\eta_\rho (f)$ for the KVORcut03 model are demonstrated in Fig. \ref{Fig-1-new} by dashed lines. The bars indicate maximum values of $f$ reachable in NSs.

 \begin{figure}
\centering
\includegraphics[width=4.5cm]{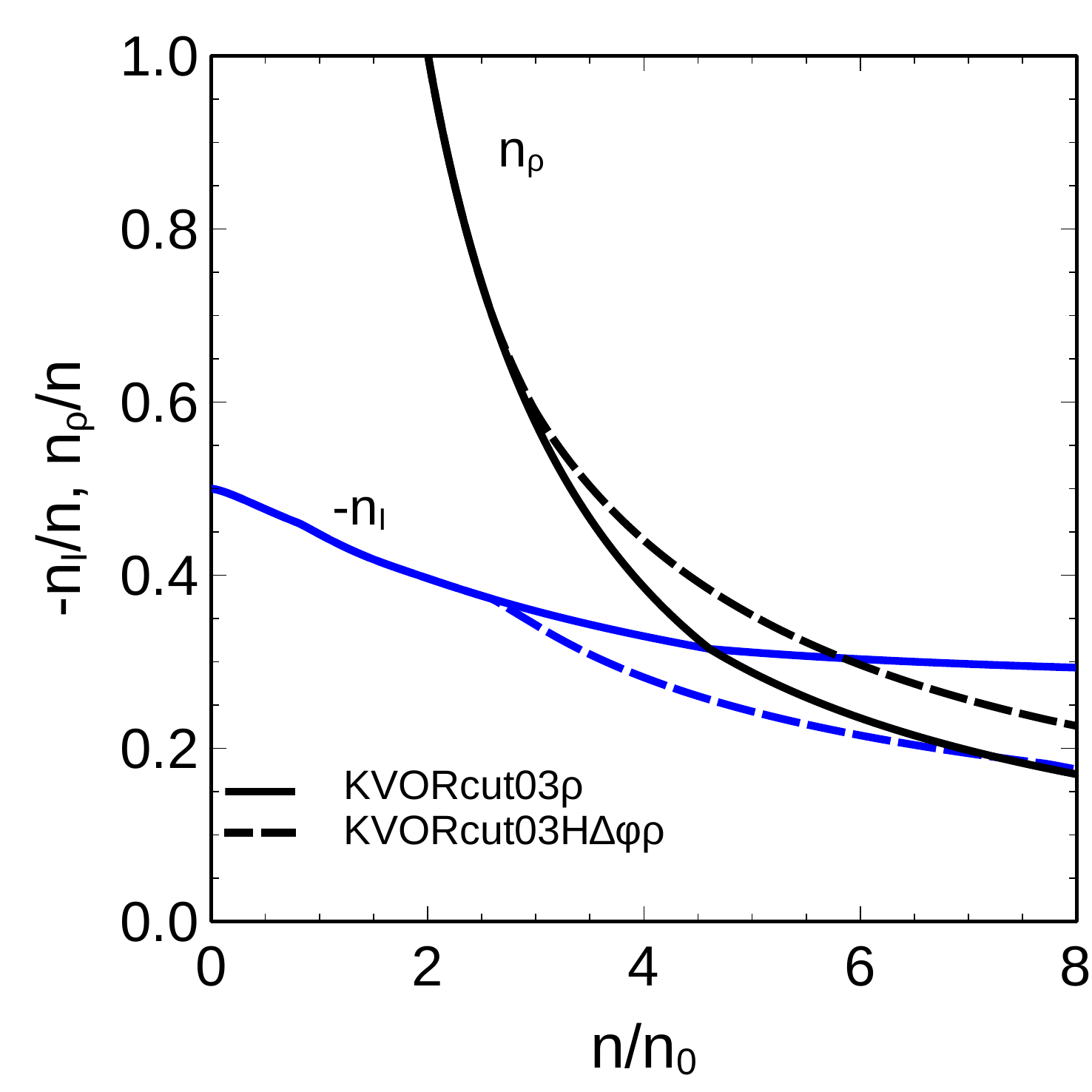} 
\includegraphics[width=4.5cm]{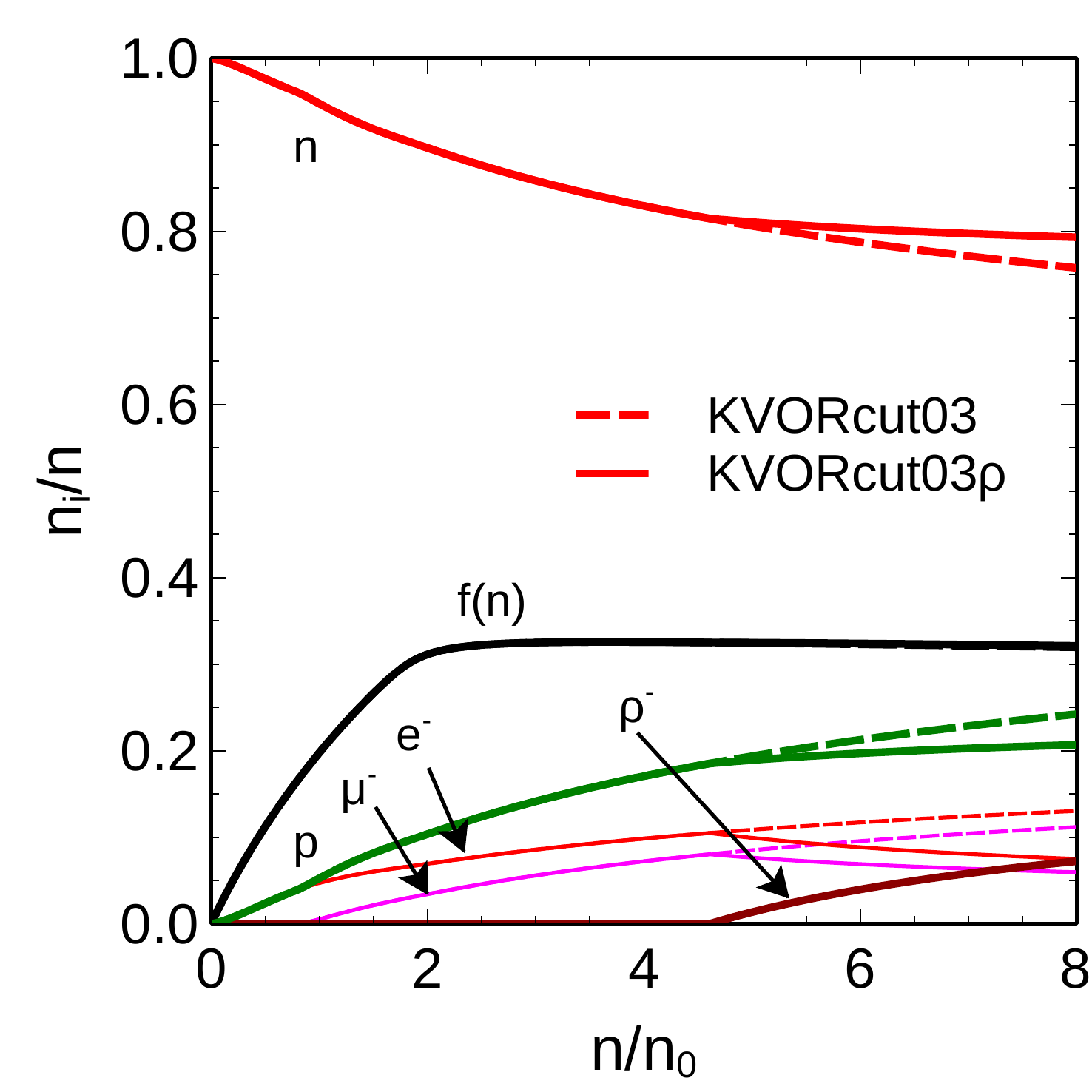} 
\includegraphics[width=4.5cm]{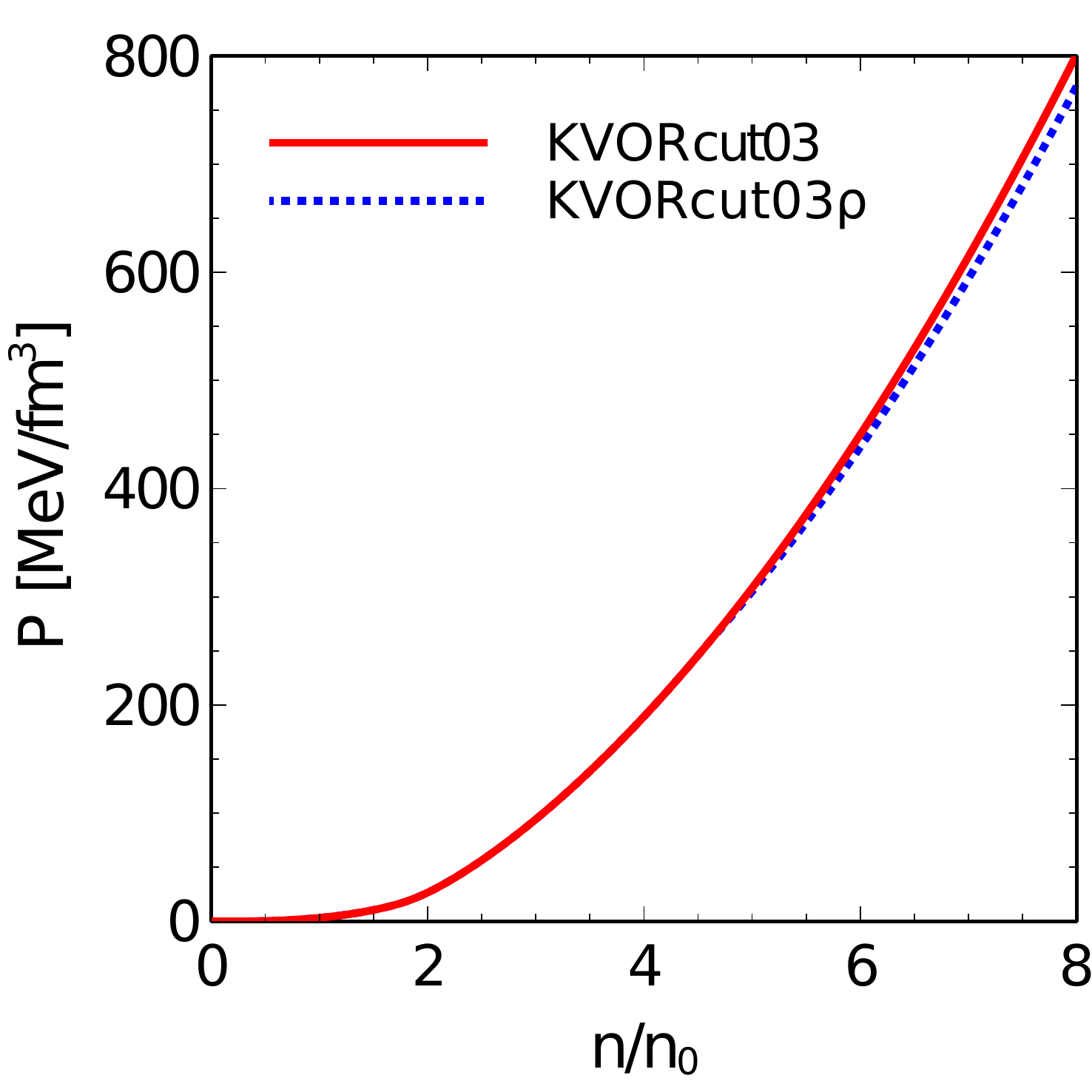} 
\caption{Left panel: the ratio of the isospin density (\ref{edensity}) to the total baryon density, $n_I/n$, and the ratio $n_\rho/n$ with $n_\rho$ from (\ref{solc})  as functions of the total baryon density in BEM for KVORcut03$\rho$ and KVORcut03H$\Delta\phi\rho$ models.
Middle panel: particle concentrations and $f(n)$ in BEM for KVORcut03 and KVORcut03$\rho$ models.
Right panel: pressure in BEM as a function of the total baryon density for the KVORcut03 and KVORcut03$\rho$ models.}
\label{fig:cut03_nr}
\end{figure}

On the left panel of Fig.~\ref{fig:cut03_nr}   for the KVORcut03$\rho$ (no hyperons, no $\Delta$s are included) and KVORcut03H$\Delta\phi\rho$ models we plot the absolute value of the isospin density $|n_I|=-n_I$ in BEM, defined in Eq.~(\ref{edensity}), and the parameter $n_\rho$ determining the amplitude of the $\rho^-$ condensate, see Eq.~(\ref{solc}), as functions of the total baryon density $n$. For the KVORcut03$\rho$ model (solid lines) $|n_I|$ exceeds $n_\rho$ for densities $n>n_{c,\rho}\simeq 4.6\,n_0$.  The presence of hyperons and $\Delta$ baryons reduces $|n_I|$. Dashed lines on the left panel of Fig.~\ref{fig:cut03_nr}  do not cross ($|n_I|<n_\rho$)  for densities $n < 8\, n_0$, so the $\rho^-$ condensation in a NS does not happen within the KVORcut03H$\Delta\phi$ model. Thus we have found that the $\rho^-$ condensate does not occur in the KVORcut03-based model at densities relevant for NSs, if hyperons and/or $\Delta$s are included.

The $\rho^-$ concentration for the KVORcut03$\rho$ model (equal to $-n_{\rm ch,\rho}/n$, where $n_{\rm ch,\rho}$ is given by Eq.~(\ref{nchrho}) for $\zeta =-1$)  as a function of the baryon density in BEM is shown on the middle panel of  Fig.~\ref{fig:cut03_nr} together with concentrations of other species and $f(n)$. We see that presence of the $\rho^-$ condensation  results in a decrease of the proton concentration and correspondingly in a decrease of lepton concentrations and in an increase of the neutron concentration.

Pressure curves for the KVORcut03 and  KVORcut03$\rho$ models as functions of the density in BEM are shown on the right panel of  Fig.~\ref{fig:cut03_nr}. We see that the $\rho^-$ condensation appears by  a second-order phase transition and the influence of the $\rho^-$ condensation on the EoS is minor. As a result, the maximum NS mass  decreases only slightly, from $2.17\, M_\odot$ to $2.16\, M_\odot$.

\subsection{MKVOR*-based models}\label{sect:MKVOR}

In MKVOR-based models a sharply varying scaling function $\eta_\rho$ is exploited in the $\rho$ sector to suppress the growth of the scalar field with an increase of the density in BEM without changing the results for ISM. To eliminate the decrease of the $\Delta$ effective mass to zero the MKVOR* extension of the model was introduced in~\cite{MKVOR-Delta}. The scaling functions $\eta_M (f)$ for the MKVOR*-based models in BEM are demonstrated in Fig.~\ref{Fig-1-new} by solid lines. Bars show maximum values of $f$ reachable in a NS.
In contrast with the KVORcut03-based models, the charged $\rho^-$ meson condensate appears in the MKVOR*-based models for any set of baryon species included in calculations. To be specific below we focus on the incorporation of a possibility of the $\rho^-$ condensation in the MKVOR*H$\Delta\phi$ model~\cite{MKVOR-Delta}. The latter includes all relevant degrees of freedom investigated in our previous works. The resulting extension will be denoted as MKVOR*H$\Delta\phi\rho$\, model.

A remark is in order.
As was mentioned already in~\cite{Kolomeitsev:2004ff}, in RMF models with the scaling functions dependent on the scalar field the equation determining the magnitude of the scalar field may have several branches of solutions for $f(n)$. When the energy for solution on the new branch  becomes lower than on the old one the system undergoes a  phase transition, provided it is considered at fixed baryon density. At a first-order phase transition the first derivative $\rmd E(n)/\rmd n$ or the energy itself change abruptly at $n=n^{(\rm I)}_c$. In our case the transition to the new branch happens with the continuous energy, but discontinuous pressure, see Fig.~\ref{fig:MKVtails-press} below.  Thus one deals with the first-order phase transition provided the system is at fixed density. The pressure as a function of the density acquires the van der Waals-like form and at a fixed pressure the first-order phase transition may occur at a smaller value of $n$,  see discussion below.
There should be a physical reason for such a phase transition otherwise the additional solutions should be considered as spurious and ought to be eliminated. As has been demonstrated in~\cite{MKVOR-Delta}, the parametrization (\ref{etarho-orig}) labeled as tail 1  (shown on the right panel of Fig.~\ref{Fig-1-new}) leads to  appearance of the new branch $f(n)$ with an  energetically favorable solution in BEM at a larger value of the $\sigma$ field than that related to old branch. Since we see no physical reason for such a phase transition to occur, we modified $\eta_\rho (f)$ to eliminate the new solution. We have suggested in~\cite{MKVOR-Delta} several modifications of the tail of the $\eta_\rho$ function for $f > f_\rho^*>f_{\rm lim}$, where $f_{\rm lim}$ is the maximal (limiting) value reachable in NSs, see parameterizations (\ref{eta-MKVOR*}), (\ref{tail123}) with labels ``tail~2, 3 and 4'' shown in Fig.~\ref{Fig-1-new}. We note here that (in difference with our previous statement in~\cite{MKVOR-Delta}) in the models MKVOR and MKVOR* without hyperons in the BEM with the tails 2,3 and 4 the second solution is not eliminated,  whereas it is eliminated for tails 2, 3 and 4 in our final MKVOR(*)H$\phi$ and MKVOR(*)H$\Delta\phi$ extensions of the model (in~\cite{MKVOR-Delta} without inclusion of the $\rho^-$ condensate). Concluding the remark, the new solutions do not appear in all our models with hyperons and/or $\Delta$s without the $\rho^-$ condensate for tails 2, 3 and 4, including the MKVOR*H$\Delta\phi$ model of our interest here.

\begin{figure}
\centering
\includegraphics[width=6.5cm]{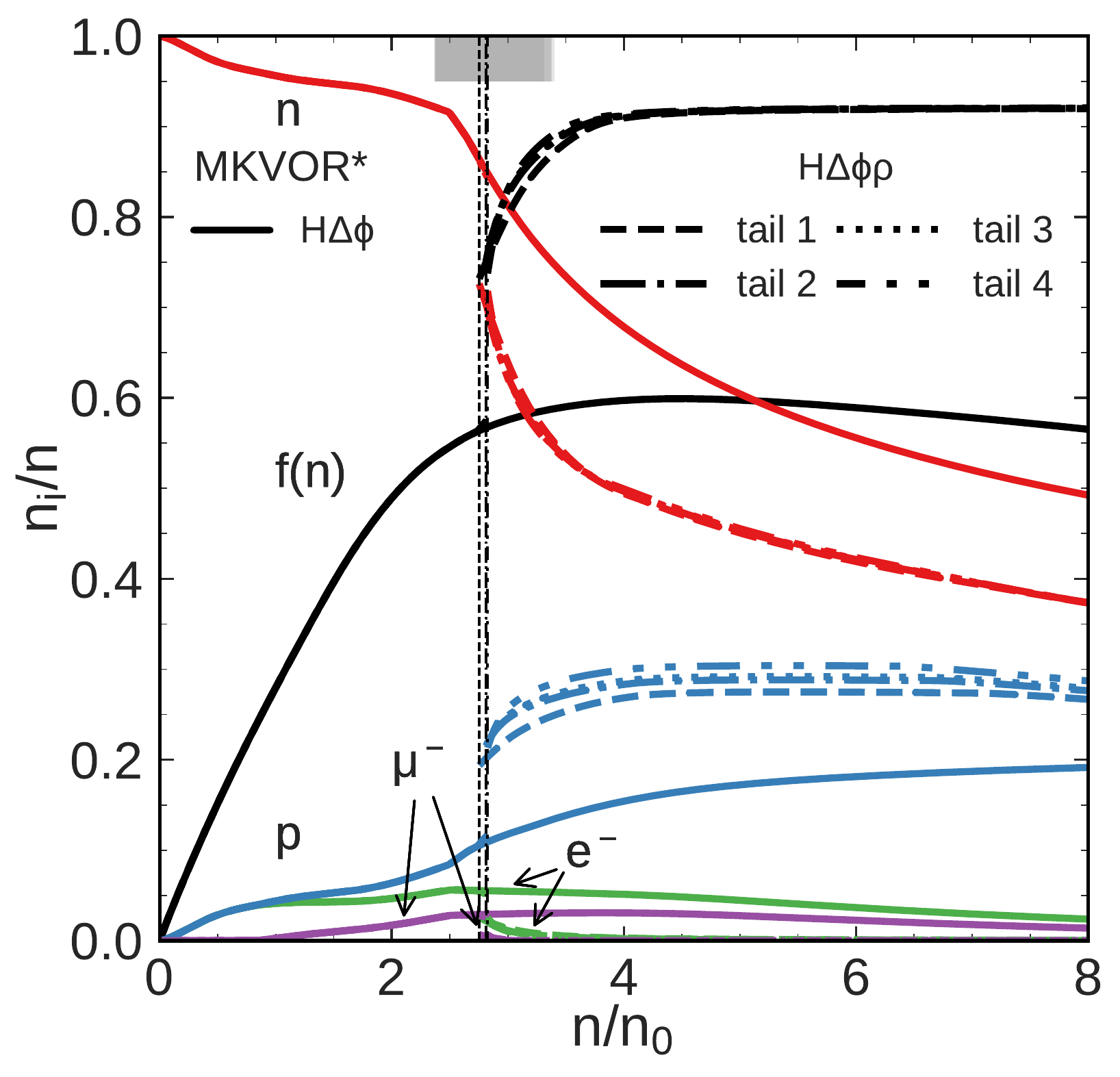}  
\includegraphics[width=6.7cm]{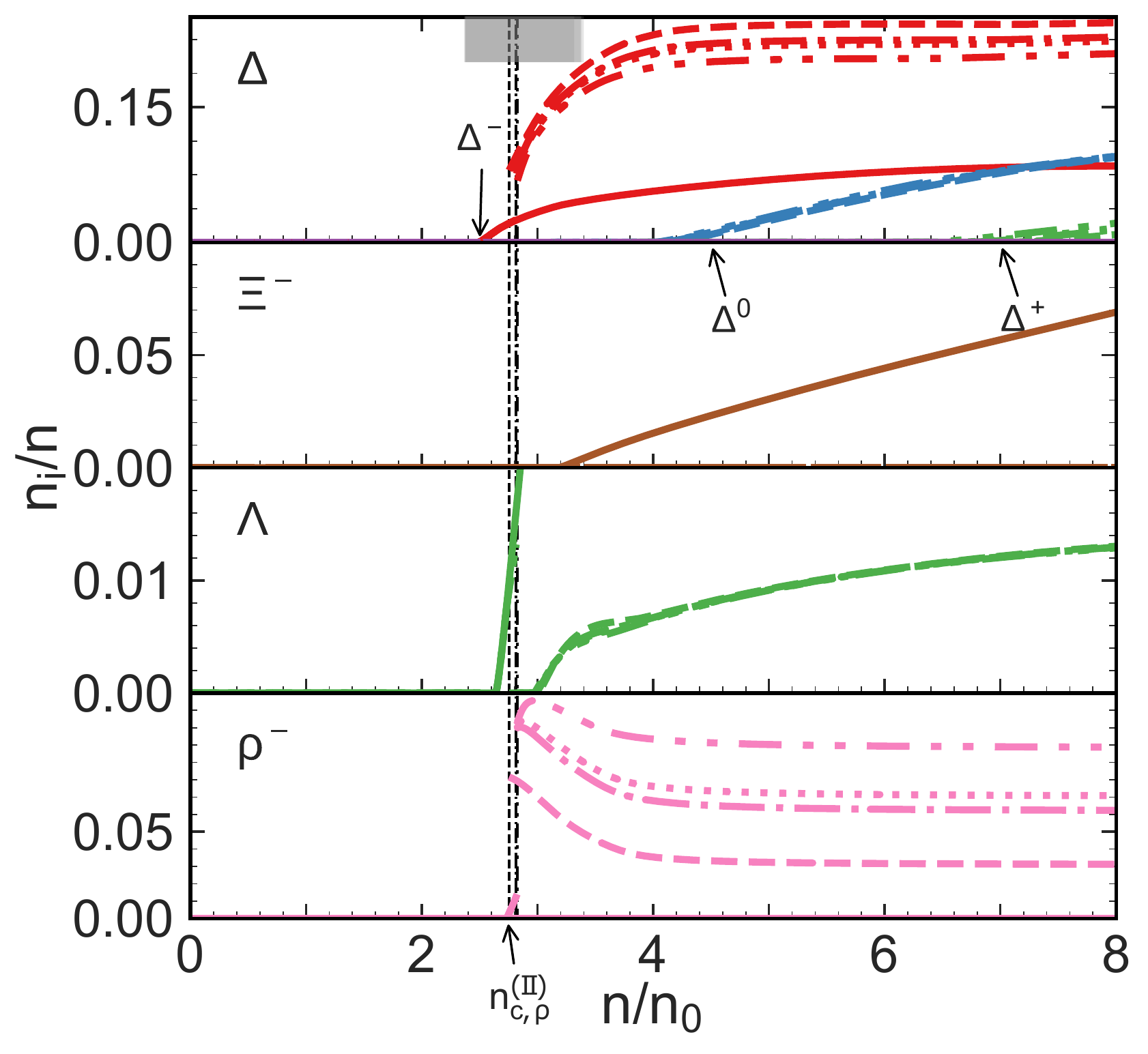}  
\caption{Left panel: proton, neutron and lepton  concentrations in BEM together with the scalar field $f$ as functions of the baryon density for the MKVOR*H$\Delta\phi$ model (solid lines), and for the MKVOR*H$\Delta\phi\rho$ model (broken lines) with various tails of the $\eta_\rho$ scaling function. Thin vertical lines denote the densities $n_{\rm c,\rho }^{\rm (I)}$ at which the  new branch of solutions becomes energetically favorable. Grey bands  show the interval of densities within the Maxwell construction. The variations of values of the initial and final densities for the Maxwell constructions for various tails are almost indistinguishable by eye here (see Fig.~\ref{fig:MKVtails-press} below for details). Right panel: the same as on the left panel but for $\Delta$,  hyperon and $\rho^-$ condensate concentrations.   The critical density of the second-order phase transition is indicated by an arrow on the $x$-axis.}

\label{fig:MKVtails-conc}
\end{figure}

The situation changes, if the $\rho^-$ condensate is taken into account. Additional branches of solution $f(n)$ in the BEM appear now for all four tails of $\eta_\rho (f)$.
In all cases the new solutions have now  clear physical meaning, describing possibility of the phase transition of the system to the $\rho^-$ condensate state. Therefore we can consider the $\rho^-$ condensation as a trigger for a jump to the new solution, provided the transition to this branch is energetically favorable.

In Fig.~\ref{fig:MKVtails-conc} (left and right)  by various broken lines we show particle fractions and the scalar field $f$  as functions of the density for the MKVOR*H$\Delta\phi\rho$ model with tails 1 through 4 for $\eta_\rho(f)$. At the density $n_{\rm c, \rho}^{\rm (II)}=2.74\,n_0$ (indicated by an arrow on the right panel of Fig. \ref{fig:MKVtails-conc}) the  appearance of the condensate by a second-order phase transition (on ``old'' branch of solutions) lowers the energy density independently on the choice of the tail.  For $n < 2.76 \, n_0$ the results for all tails in MKVOR*H$\Delta\phi\rho$ model coincide among each other. However, with the density increase another branch (``new" branch) of solutions becomes energetically favorable at densities $n_{\rm c,\rho }^{\rm (I)}\simeq (2.76,2.81,2.82,2.83) \, n_0$ for tails 1, 2, 3 and 4, respectively\footnote{Note that in the MKVORH$\Delta\phi$ model (without inclusion of the $\rho^-$ condensate) for tail 1  the second (``spurious") solution appears at a higher density $n=3.34\,n_0$.}. The concentration of $\rho^-$ on the ``old" branch don't exceed $0.02$ before the transition to another branch, and can barely be seen on the Fig. \ref{fig:MKVtails-conc}. The new branch is featured by a larger fraction of $\rho^-$ condensate. The densities $n_{\rm c, \rho}^{\rm (I)}$ are shown in Fig. \ref{fig:MKVtails-conc} by thin vertical lines, with the line styles corresponding to the models as in the legend. We show only the energetically favorable parts of the solution branches. By the solid lines on both panels of Fig. \ref{fig:MKVtails-conc} we show the solution for $f(n)$ and particle fractions for MKVOR*H$\Delta\phi$ model without $\rho^-$ condensation for comparison. Curves for all tails coincide with solid ones for $n < n_{\rm c, \rho}^{\rm (II)}$.

The new particle species appear before the condensate appears, cf. Table \ref{tab:MKVOR-crit-param}. The neutron density decrease rate changes at $n \simeq 2.6 \,n_0$, which is a consequence of the almost simultaneous appearance of $\Delta^-$ and $\Lambda$-baryons. At $n=n_{\rm c,\rho }^{\rm (I)}$ the equilibrium state of the system for fixed total baryon density switches to the new branch, resulting in an abrupt increase of the amplitude of the $\rho^-$ condensate from tiny values corresponding to the old branch to the finite ones on the new branch. The increase of the condensate amplitude leads to abrupt changes in concentrations of neutrons, protons, electrons and $\Delta^-$ and to the disappearance of $\Lambda$ hyperons. The electron and muon fractions (curves indicated by arrows on the left panel of Fig. \ref{fig:MKVtails-conc})  visually coincide among all tails for $n > n_{\rm c, \rho}^{\rm (I)}$, and they disappear completely at $n \simeq 3 \, n_0$ and $4 \, n_0$, respectively. The leptons are partially  replaced by $\Delta^-$s, which concentration increases by a factor larger than 2 (see on right panel). The proton fraction increases also by 50\%. The proton, $\Delta^-$ and $\rho^-$  concentrations demonstrate  strong dependence on the choice of the tail. Despite the notable variation of the  concentrations of charged species, the neutron, $\Lambda$ and $\Delta^0$ fractions prove to be almost independent on the choice of the tail. Gray bands in Fig.~\ref{fig:MKVtails-conc} mark the intervals of densities bridged over by Maxwell constructions (see further discussion of  Fig.~\ref{fig:MKVtails-press}). Such densities are not realized in equilibrium configurations of NSs.

From Figs.~\ref{Fig-1-new} and~\ref{fig:MKVtails-conc} we conclude that the lower  tail of $\eta_\rho(f)$ for $f>f_\rho^*$  is chosen (tail 4 is the lowest one) the higher is the amplitude of the $\rho^-$ condensate and the proton concentration, but the smaller is the $\Delta^-$ concentration. In contrast to the MKVOR*H$\Delta\phi$ model, in MKVOR*H$\Delta\phi\rho$ model $\Delta^0$ and $\Delta^+$ baryons do appear in the medium. The corresponding critical densities can be found in Table~\ref{tab:MKVOR-crit-param}. They depend on the choice of a tail and are varied within the interval $(4.07 - 4.20) \, n_0$ for $\Delta^0$, and $(6.5 - 7.23)\, n_0$ for $\Delta^+$. The $\Xi^-$ hyperons do not appear in the model with $\rho^-$ condensate.
\
\begin{figure}
\centering
\includegraphics[width=6.5cm]{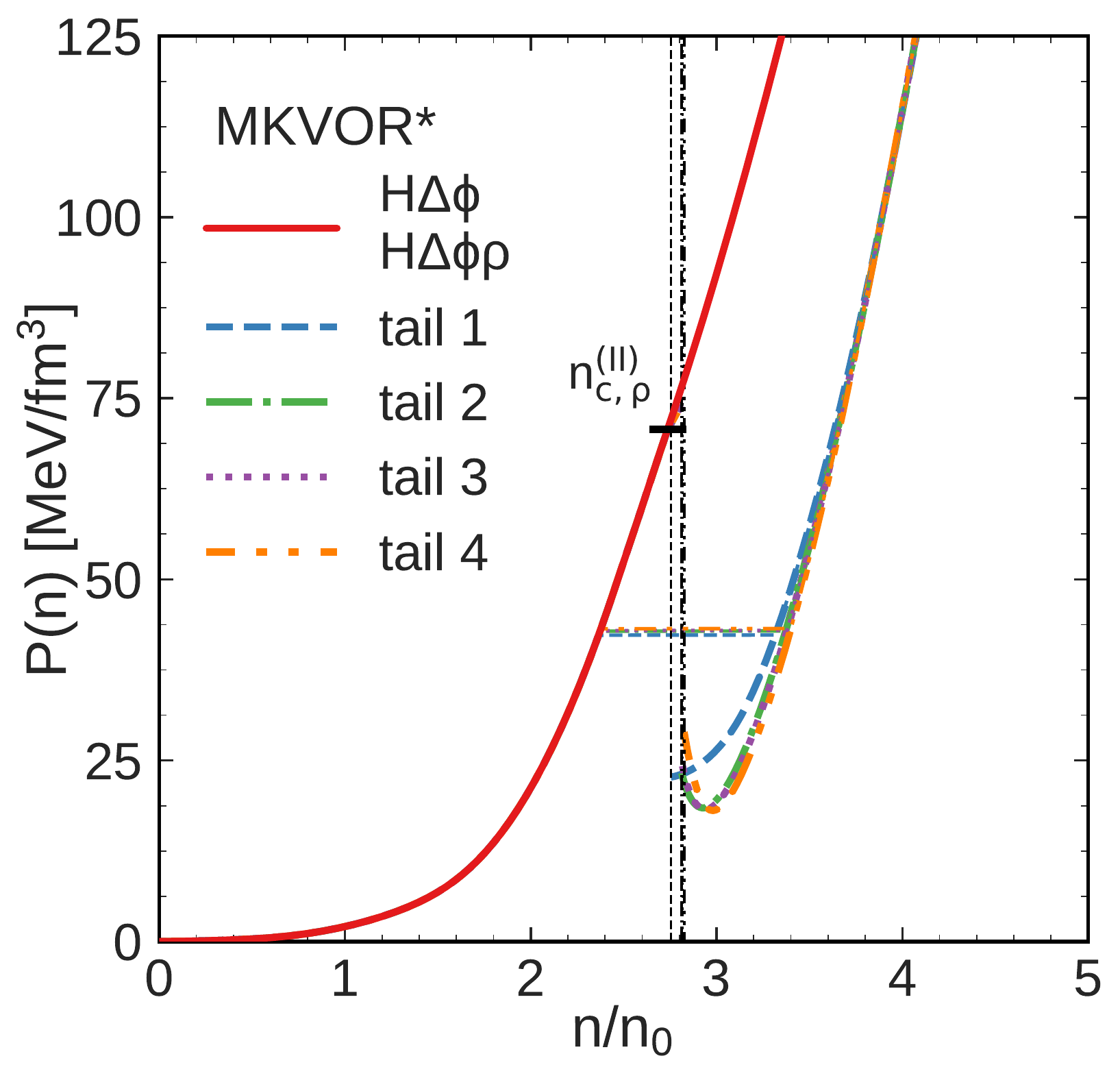} 
\includegraphics[width=6.5cm]{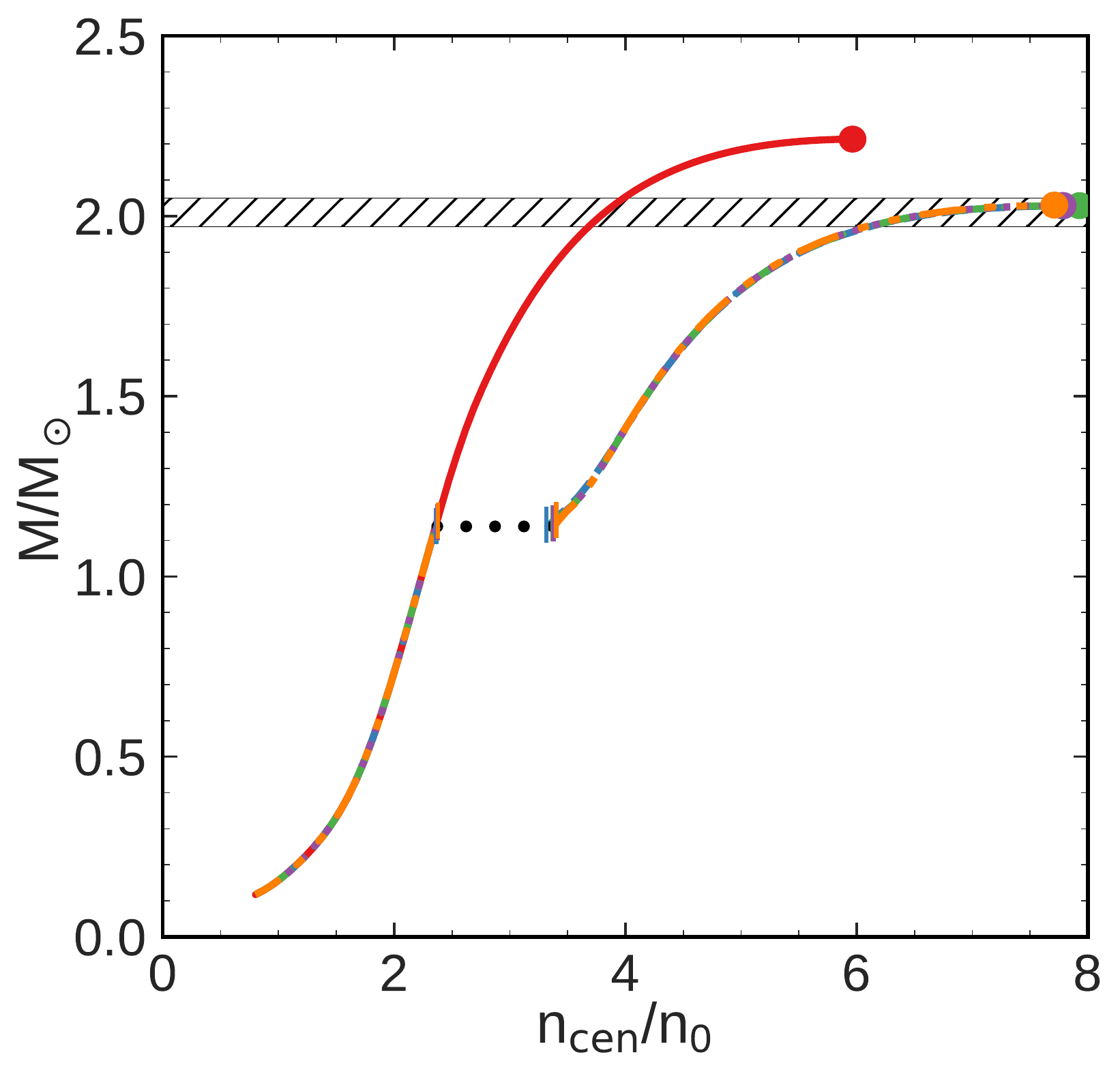} 
\caption{Left panel: pressure of BEM as a function of the density with Maxwell construction shown by thin horizontal lines.  Solid lines are for the MKVOR*H$\Delta\phi$ model
and four types of other lines show the results for
the MKVOR*H$\Delta\phi\rho$ model with four tails of $\eta_\rho(f)$.
The horizontal bar on the left panel indicates the critical density, $n_{\rm c,\rho}^{\rm (II)}$, of the appearance of the $\rho^-$ condensate by a second-order phase transition, and vertical lines mark the densities $n_{\rm c,\rho}^{\rm (I)}$.  Right panel: NS mass as a function of the central density. The hatched region  shows the maximum measured mass of the pulsar PSR~J0348+0432, $M = 2.01 \pm 0.04 \, M_\odot$~\cite{Antoniadis:2013pzd}.}
\label{fig:MKVtails-press}
\end{figure}

The transition between solutions we described above is an example of a first-order phase transition, if the system is considered at a fixed density. However, in stellar interiors a proper thermodynamical variable is the local pressure, whose continuity along the radial profile is required by the mechanical stability condition (one of the Gibbs conditions). On the left panel in Fig.~\ref{fig:MKVtails-press} we show the pressure in the BEM as a function of the density for MKVOR*H$\Delta\phi$ and MKVOR*H$\Delta\phi\rho$ models. For the MKVOR*H$\Delta\phi$ model $P(n)$ demonstrates a monotonous behavior (solid line).
The  $\rho^-$ condensation occurs by the first-order phase transition. The transition to the new branch $f(n)$ for $n>n_{\rm c,\rho}^{\rm (I)}$ in the MKVOR*H$\Delta\phi\rho$ model leads to the discontinuity in the $P(n)$ curve (see broken lines). Similar behavior has been  studied in detail in ~\cite{MKVOR-Delta}, cf.  Fig.~12 there.  Thin horizontal lines indicate equilibrium configurations corresponding to the Maxwell constructions.
The Maxwell-construction lines start at the densities $n^{\rm (MC)}_1 \simeq 2.37\,n_0$ for tails 1,2, and 3 and at $n^{\rm (MC)}_1 \simeq 2.38\,n_0$ for tail 4 and end at
$n^{\rm (MC)}_2=(3.32$, 3.37, 3.38, $3.40)\,n_0$ for tails 1--4, respectively. We see that the pressure depends significantly  on the choice of the tail only for densities $n_{\rm c,\rho}^{(\rm I)}<n< n_2^{\rm (MC)}$ and
is almost tail-independent for  $n>n_2^{\rm (MC)}$ despite
the matter composition changes strongly with a variation of the $\eta_\rho$ tail. Along the Maxwell-construction line the matter is in mixed phase with an averaged density, which varies between $n^{\rm (MC)}_1$ and $n^{\rm (MC)}_2$ according to the equation $\bar{n}= n^{\rm (MC)}_1 (1-f_{\rm \rho c})+ n^{\rm (MC)}_2 f_{\rm \rho c}$, where $f_{\rm \rho c}$ is the relative fraction of the volume occupied by the $\rho^-$ condensate phase.
Thus already when the density in the center of a NS reaches the value $n_1^{\rm (MC)}$, after a while there appears a droplet of the $\rho^-$ condensate phase with $n=n^{\rm (MC)}_2$. In dynamics, the appearance of the new phase looks more peculiarly. Here we avoid this discussion being more relevant for heavy-ion collisions~\cite{Schulz:1983pz,Skokov:2009yu,Skokov:2010dd} and for an initial stage of the NS cooling~\cite{Voskresensky:1987ut,Haubold:1988uu,Migdal:1990vm}, and consider only equilibrium configurations.

Moreover, for the systems with one (baryon) charge conservation the equilibrium state for a given pressure $P$ should be maximal as a function of the chemical potential $\mu$.
This is equivalent to using of the Maxwell construction ($P=const$ for $n_1^{\rm (MC)}<n<n_2^{\rm (MC)}$)  describing  equilibrium state, as shown in left panel of Fig.~\ref{fig:MKVtails-press}. In a NS besides the baryon number, the electric and lepton charges are conserved. The electric charge may be conserved globally rather than locally. This leads to a possibility of occurrence of the pasta phase with the pressure changing continuously with the averaged density $\bar{n}$, cf.~\cite{Glend92}. Following results of various studies, cf.~\cite{Glendenning}, pasta phase  does not substantially influence the EoS although the particle concentrations change significantly. Besides, the mixed phase appears only, if the surface tension between two phases is below a critical value~\cite{Heiselberg,Voskresensky:2001jq,Voskresensky:2002hu}.  Finite-size Coulomb and surface effects result in that for realistic values of the surface tension the pressure in the range of the pasta phase is close to that for the Maxwell construction~\cite{Voskresensky:2002hu,Maruyama:2005tb}. Thus, if mixed phase occurs, it should not change essentially  the maximum NS mass, which is the most important quantity we address in this work. Thereby, below we continue to consider a  simpler case of the Maxwell construction.

Note that a strong first-order phase transition of a NS to the $\rho^-$ condensate state can be manifested in such a phenomenon as blowing-off of a part of a star accompanied by a neutrino burst. For the strong first-order phase transition in a NS to a state with a pion condensate these phenomena were discussed in~\cite{Voskresensky:1987ut,Haubold:1988uu,Migdal:1990vm}. Similar phenomena may occur in case of the charged $\rho$ condensation. So, if a strong first-order phase transition to the $\rho^-$ condensate state occurred during formation or cooling of the newly born NS in supernova explosion it may be manifested through the second neutrino burst delayed typically by $t\sim 10$\,s (may be  up to $t\sim$ several hours) compared to the first one. For a rather low mass cold NS in a binary system, the transition may occur through accumulation of the matter in the accretion process.

\begin{table}
\caption{Critical densities and NS masses for appearance of various fractions and those for DU processes on neutrons, maximum central densities and NS masses, and corresponding radii, and radii for the NS with the mass $1.5\,M_\odot$ for MKVOR*H$\Delta\phi$ and MKVOR*H$\Delta\phi\rho$ models. For the MKVOR*H$\Delta\phi$ model  results do not depend on the choice of the tail in $\eta_\rho$. Long dashes denote that  the species do not appear in the NS within these models. Asterisks mark  densities within the Maxwell construction area, therefore, corresponding to the same NS mass of 1.14\,$M_\odot$.
}
\centering
\begin{tabular}{|clc|c|c|c|c|}
\hline
& & MKVOR*H$\Delta\phi$ &\multicolumn{4}{|c|}{MKVOR*H$\Delta\phi\rho$} \\\hline
 &      &               &  tail 1 & tail 2  & tail 3 &  tail 4 \\
\hline
$n_{ c\rho}^{\rm I}$ &[$n_0$]     &---        & 2.76* & 2.81* & 2.82* & 2.83* \\\cline{4-7}
$n_{ c\rho}^{\rm II}$ &[$n_0$]     &---        & \multicolumn{4}{|c|}{2.74*} \\ \cline{4-7}
$n_{1}^{\rm MC}$ &[$n_0$]     &---        & 2.37 & 2.37 & 2.37 & 2.38 \\
$n_{2}^{\rm MC}$ &[$n_0$]     &---        & 3.32 & 3.37 & 3.38 & 3.40 \\
\hline\hline
$n_{{\rm c},\Delta^-}$ &[$n_0$] & 2.51 & \multicolumn{4}{|c|}{2.51*} \\ \cline{4-7}
$M(n_{{\rm c},\Delta^-})$ &[$M_\odot$] & 1.30 & \multicolumn{4}{|c|}{1.14\phantom{*}} \\ \cline{4-7}

$n_{{\rm c},\Lambda}$ &[$n_0$] & 2.66 & \multicolumn{4}{|c|}{2.66*} \\ \cline{4-7}
$M(n_{{\rm c},\Lambda})$ &[$M_\odot$] & 1.44 & \multicolumn{4}{|c|}{1.14\phantom{*}} \\  \cline{4-7}

$n_{{\rm c},\Delta^0}$ &[$n_0$] &---  &4.20 &4.13 &4.12  &4.07 \\
$M(n_{{\rm c},\Delta^0})$ &[$M_\odot$] &---        & 1.51 & 1.48 & 1.47 & 1.44 \\
$n_{{\rm c},\Delta^+}$ &[$n_0$]  &---        &7.23 & 6.88 &6.79 &6.50 \\
$M(n_{{\rm c},\Delta^+})$ &[$M_\odot$] &---  & 2.02 &  2.02 & 2.01&2.00 \\ \cline{4-7}
$n_{{\rm c},\Xi^-}$ &[$n_0$]   &3.24  & \multicolumn{4}{|c|}{------} \\ \hline
\hline
$n_{\rm c,DU}^n$ &[$n_0$]         & 3.60 & --- &5.64 & 5.32 & 4.51 \\
$M_{\rm DU}^n$ &[$M_\odot$]       & 1.94 & --- &1.91 & 1.87 & 1.65 \\
\hline\hline
$ n_{\rm cen}^{\rm max}$ & $[n_0]$      & 6.03 & 7.64 &7.50 & 7.64 & 7.86 \\
$M_{\rm max}$ & [$M_\odot$]      & 2.21 & 2.03 &2.03 & 2.03 & 2.03 \\
$R[M_{\rm max}]$ &[km]          & 11.11& 9.78 &9.82 & 9.78 & 9.72 \\
$R(1.5\,M_\odot)$ &[km]         & 12.14& 11.10 &11.12& 11.12 & 11.13 \\
\hline\hline
\end{tabular}
\label{tab:MKVOR-crit-param}
\end{table}
On the right panel of Fig.~\ref{fig:MKVtails-press}  we show the NS mass for the MKVOR*H$\Delta\phi$ and MKVOR*H$\Delta\phi\rho$ models as a function of the central density. The resulting NS mass proves to be almost independent on the $\eta_\rho$ tail, with the maximum value $M_{\rm max} \simeq 2.03 \, M_\odot$ for MKVOR*H$\Delta\phi\rho$ model. Thus although the $\rho^-$ condensation leads to a reduction of the maximum NS mass by $\sim 0.2\,M_\odot$, the resulting MKVOR*H$\Delta\phi\rho$ model is still compatible with empirical constraints~\cite{Demorest:2010bx,Fonseca,Antoniadis:2013pzd}.
The radius of the maximum-mass NS reduces after the inclusion of the $\rho^-$ condensate by 1.3--1.4~km, see Table~\ref{tab:MKVOR-crit-param}. The star of a typical mass 1.5\,$M_\odot$ also becomes  more compact in presence of the $\rho^-$ condensation,  the radius decreases from $\approx$ 12.1~km to $\approx$ 11.1~km. All the global star properties  depend very weakly on the choice of the tail of the $\eta_\rho$ scaling function.

The critical density of the direct Urca (DU) reactions on nucleons, $n_{\rm c,DU}^n$, increases with inclusion of the $\rho^-$ condensate due to disappearance of the leptons, and the corresponding NS mass is sufficiently high for all the tails, thereby the model satisfies the constraint $M_{\rm DU}^n >1.5\,M_\odot$, see Table~\ref{tab:MKVOR-crit-param}. The critical density for appearance of $\Delta^-$ is rather low, $2.51 n_0$ (being within the Maxwell construction) corresponding to the NS mass $1.14 M_{\odot}$.
However the DU reactions on $\Delta$s may occur only when the DU processes on nucleons are allowed~\cite{Prakash-DU}, and therefore they do not spoil the DU constraint~\cite{Kolomeitsev:2004ff,Klahn:2006ir,Maslov:2015wba}.
The critical density of the  appearance of $\Lambda$s is also within the Maxwell construction,  $n_{\rm c,\Lambda}=2.66 n_0 < n_2^{\rm (MC)}$, and hence any star with the mass $M>M(n_1^{\rm MC})=M(n_2^{\rm (MC)})=1.14\,M_\odot$ would contain a fraction of dense matter $(n\ge n_2^{\rm (MC)})$ with $\Lambda$ hyperons. The DU reactions on $\Lambda$s are allowed even for a tiny amount of $\Lambda$s in the matter \cite{Prakash-DU}, so they can be operative in NSs with masses as low as $1.14\, M_\odot$. Although the emissivity of the neutrino DU processes $\Lambda \to p+e+\bar{\nu}_e$, $p+e\to \Lambda +\nu_e$  is suppressed by a factor $\sin^2 \theta_{\rm C}\simeq 0.05$ compared to that for the DU processes on nucleons, the presence of  DU reactions on $\Lambda$s may cause some troubles with the description of NS cooling within the   MKVOR*H$\Delta\phi\rho$ model. The problem could be avoided, if the $\Lambda$ Cooper pairing gaps were sufficiently large, cf. \cite{Takatsuka:2005bp}. We will return to this question in a subsequent publication. Also, in the presence of the $\rho^-$ condensate  new  neutrino-emission processes become possible, e.g., $n+\rho^-_c \to n +e+\bar{\nu}_e$, the emissivity of this process
was estimated in~\cite{Kolomeitsev:2004ff} to be of the order either smaller than that for the processes on the charged pion condensate \cite{Voskresensky:1986af}. Since processes on pion condensates do not spoil a general appropriate description of the NS cooling, cf.  ~\cite{Blaschke:2004vq,Grigorian:2016leu}, presence of the processes on $\rho^-$ condensate should not cause  the problems as well.

\section{Variations of $m_\rho^*(f)$ and $\eta_\rho (f)$ in the MKVOR*-based models}\label{variat}

In this section we investigate sensitivity of the $\rho^-$ condensation effect to the variations of the scaling functions for the effective $\rho$ meson mass, $\Phi_\rho(f)$, and the scaling function $\eta_\rho(f)$.

\subsection{Variation of $m_\rho^*(f)$}\label{sec:mrho}

\begin{figure}
	\includegraphics[width=\textwidth]{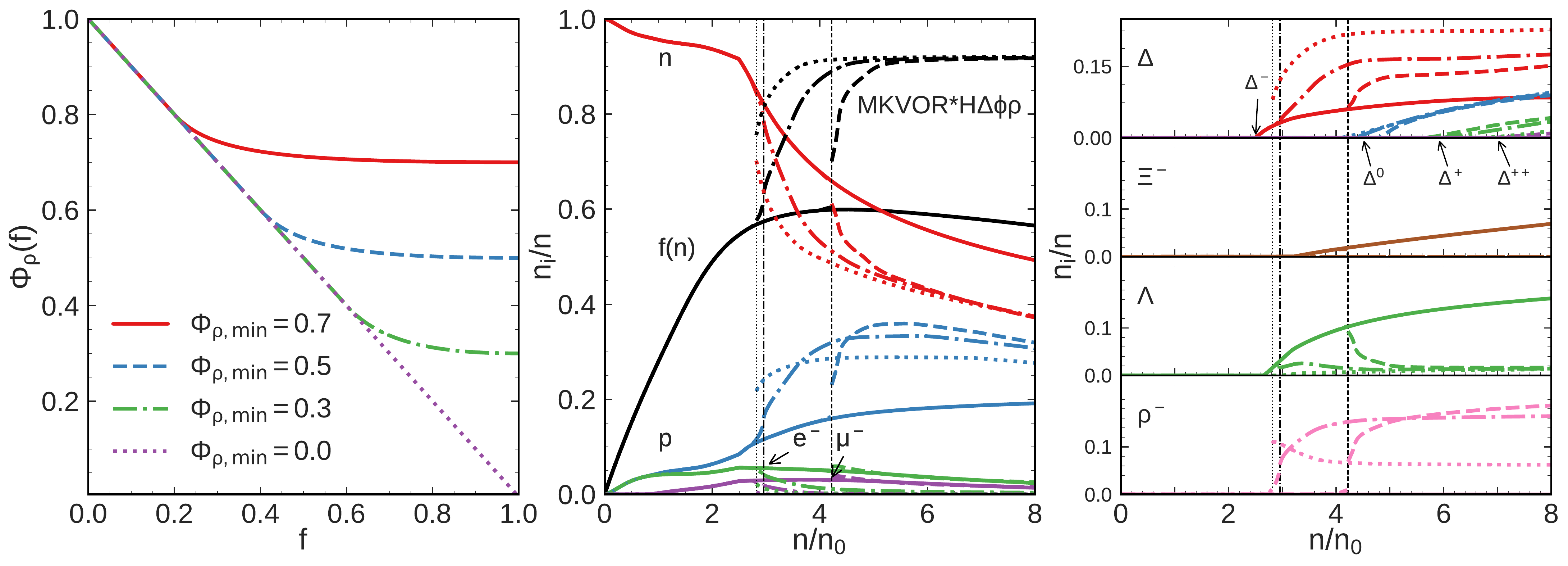} 
	\caption{
Left panel: $\Phi_\rho(f)$ for the model MKVOR*H$\Delta\phi\rho$  in BEM  with  the effective  $\rho$ meson mass scaling $\Phi_\rho(f)$ modified following Eq. (\ref{app:Phir}), with $\Phi_{\rho,\rm min} =0.3, 0.5, 0.7$,  $\eta_\rho (f)$ is taken for the tail 2. The dotted line labeled as $\Phi_{\rho,\rm min}=0$ corresponds to  $\Phi_\rho(f) = 1-f$. Middle panel: scalar field $f(n)$ together with concentrations of neutrons, protons and leptons as functions of the density  in BEM for the model MKVOR*H$\Delta\phi\rho$ for the same
 $\Phi_\rho(f)$ as in left panel.  Right panel: concentrations of $\rho^-$, $\Lambda$, $\Xi^-$ hyperons and $\Delta$s  for the MKVOR*H$\Delta\phi\rho$ model. Vertical lines on the middle and right panels denote the densities $n_{\rm c,\rho}^{\rm (I)}$.
}
	\label{fig:meff_conc}
\end{figure}

A substantial decrease of $\rho$ meson effective mass in NS matter is the necessary condition for the appearance of $\rho^-$ condensate in NS interiors.
In our approach it is modeled by the $\rho$ meson effective mass scaling function $\Phi_\rho(f)=m_\rho^*/m_\rho$ dependent on the scalar field $f$.  In the   models we have considered above, $f(n)$ grows and $\Phi_\rho(f)$ decreases with increasing density. As we have demonstrated in the previous section, with the choice of the universal scaling functions for vector mesons and nucleons $\Phi_m (f)=\Phi_N(f) = 1-f$ used in our previous studies~\cite{Maslov:2015msa,Maslov:2015wba,MKVOR-Delta} in MKVOR*-based models there is a strong first-order phase transition to the $\rho^-$ condensate state. It results in a substantial decrease of the maximum NS mass. Now let us consider another possibility, when the scaling  $\Phi_\rho (f)$ is modified such as $\Phi_\rho (f\to 1)\to \Phi_{\rho,\rm min}>0$. To be specific we  will utilize tail 2 of the $\eta_\rho$ scaling function.
We consider three cases  corresponding to $\Phi_{\rho,\rm min}=0.3$, 0.5, and 0.7, as shown on the left panel in Fig.~\ref{fig:meff_conc}. Analytical expression for the $\rho$-meson  effective-mass scaling function is given by Eq. (\ref{app:Phir}) of Appendix C. The curve marked as $\Phi_{\rho,\rm min}=0$ corresponds to the  scaling $\Phi_\rho (f)= 1-f$ we have used in previous sections.

On the middle and right panels of Fig.~\ref{fig:meff_conc} we demonstrate  $f(n)$ and particle concentrations as functions of the total baryon density. Limiting of the dependence $\Phi_{\rho}(f)$ from below  by a value $\Phi_{\rho,\rm min}$ leads to an increase of the critical density of the $\rho^-$ condensation first-order phase transition, shown by thin vertical lines for each model.
For $\Phi_\rho(f)$ with $\Phi_{\rho,\rm min}=0.3$ we find  $n_{\rm c, \rho}^{\rm (I)}=2.95\, n_0$ and with $\Phi_{\rho,\rm min}=0.5$ we have $n_{\rm c,\rho}^{\rm (I)}=4.22\, n_0$, whereas with  the original scaling $\Phi_{\rho}(f)=1-f$ the value $n_{\rm  c,\rho}^{\rm (I)}$ is $2.81\,n_0$. For $\Phi_{\rho,\rm min}=0.7$ the $\rho^-$ condensation does not occur in a NS since the new branch of $f(n)$ does not become energetically favorable up to $n  =8\,n_0$ and on the old branch the $\rho^-$ condensate does not appear at all. From the Fig.~\ref{fig:meff_conc} (right panel)
we see that although the critical density $n_{\rm c, \rho}^{\rm (I)}$ increases with an increase of $\Phi_{\rho,\rm min}$, the fraction of $\rho^-$  increases. For all $\Phi_{\rho,{\rm min}}$ the density of $\Delta^-$ appearance is $n_{\rm c}^{(\Delta^-)} = 2.51 \, n_0 < n_{\rm c,\rho}^{(\rm I)}$, which results in a rapid change in neutron and proton fractions seen in the middle panel of Fig. \ref{fig:meff_conc}. The maximum $\Delta^-$ concentration becomes the smaller, the larger the $\rho^-$ density is. The same interplay between $\Delta^-$ and $\rho^-$ fractions was noticed already in Fig.~\ref{fig:MKVtails-conc} for different tails in the original MKVOR*H$\Delta\phi\rho$ model. The increase of the $\rho^-$ condensate amplitude is accompanied by an increase of the proton fraction and a strong reduction
of the $\Lambda$ concentration (which approaches 1\%---2\% for $n \simeq 8 \, n_0$).
We see also that for $\Phi_{\rho,{\rm min}} = 0, 0.3, 0.5$ (i.e. in models, where $\rho^-$ condensation does occur) the $\Xi^-$-hyperons disappear completely immediately after the density $n_{\rm c, \rho}^{(\rm I)}$ is reached.

\begin{figure}
	\centering
	\includegraphics[width=0.32\textwidth]{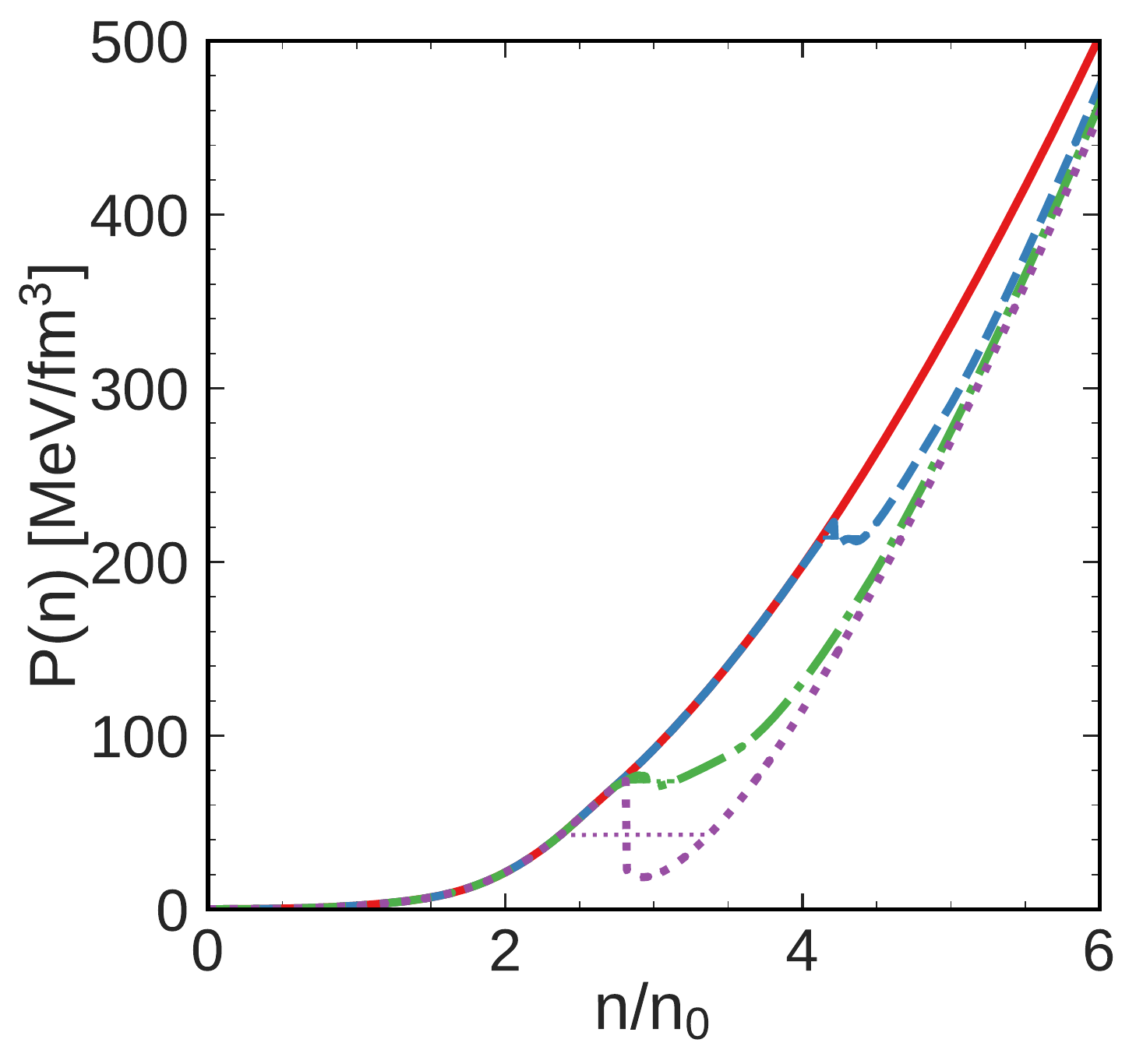} 
	\includegraphics[width=0.32\textwidth]{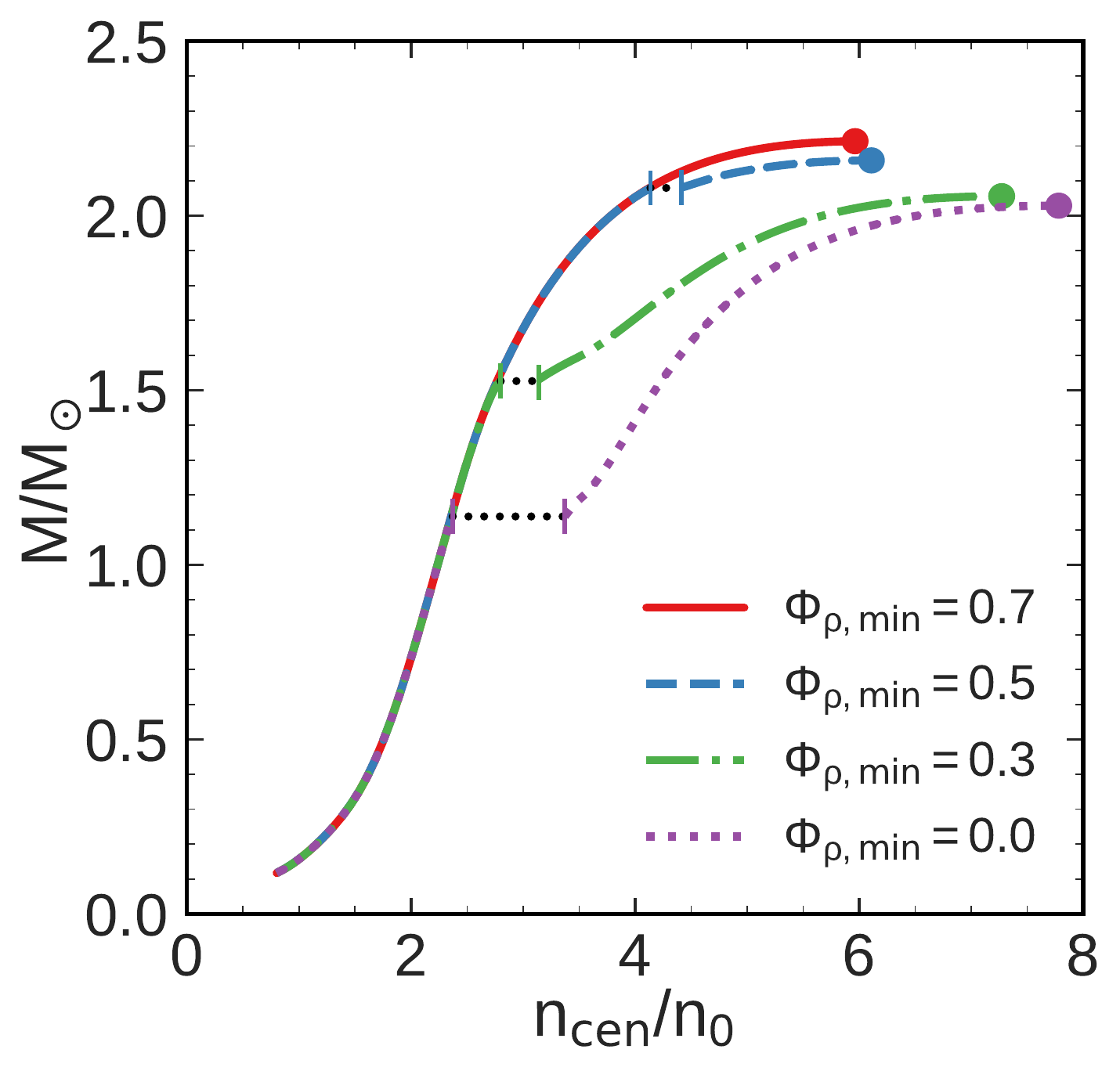} 
	\includegraphics[width=0.32\textwidth]{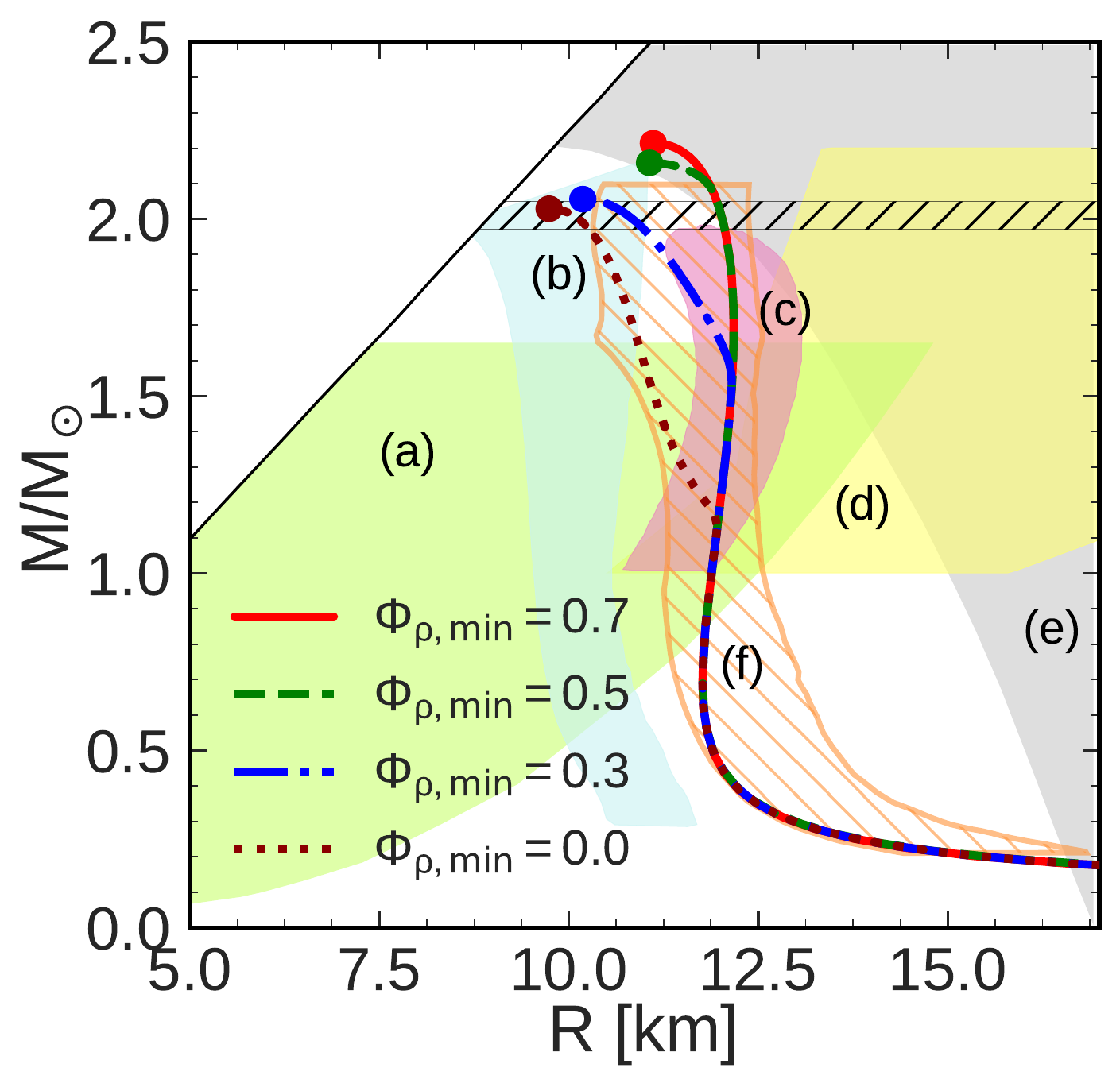} 
	\caption{Left panel: pressure as a function of the density for the MKVOR*H$\Delta\phi\rho$ model with $\Phi_\rho(f)$ given by Eq. (\ref{app:Phir}) with $\Phi_{\rho,\rm min} = 0.3$, 0.5, and 0.7. The lines denoted as $\Phi_{\rho,\rm min}=0$ are calculated with $\Phi_\rho(f) = 1-f$. Maxwell constructions are shown by thin lines. Middle panel: NS mass as a function of central density for the same models. Vertical dashes correspond to the lowest and the highest densities of the mixed phase, described by the Maxwell constructions. Neutron star configurations with the intermediate central densities cannot be realized. Blobs at the end of the lines indicate the maximum NS masses. Right panel: mass-radius relation for the same models in comparison with emprical constraints: (a)~\cite{Straaten}, (b)~\cite{Ozel:2015fia}, (c)~\cite{Suleimanov:2016llr},
(d)~\cite{Bogdanov:2012md}, (e)~~\cite{Trumper}, (f)~\cite{Lattimer:2012nd,Lattimer:2013hma,Steiner:2015aea}. The horizontal band shows the uncertainty range  for the mass of pulsar
J0348+0432~\cite{Antoniadis:2013pzd}.
}
	\label{fig:VarMeff:PnMnMR}
\end{figure}

On the left panel of Fig.~\ref{fig:VarMeff:PnMnMR} we show the pressure in the BEM for models with various $\Phi_\rho(f)$ scaling functions we consider in this section. We see that the softening of the EoS induced by formation of the $\rho^-$ condensate is the strongest for the universal scaling, $\Phi_\rho(f)=1-f$, and decreases with an increase of the value $\Phi_{\rho,\rm min}$, whereas  $n_{c,\rho}^{(\rm I)}$ increases. With $\Phi_{\rho,\rm min}=0.7$ the pressure $P(n)$ increases monotonously with the density.
Simultaneously, the density jump on the Maxwell construction, $n_2^{\rm (MC)}-n_1^{\rm (MC)}$, decreases with an increase of $\Phi_{\rho,\rm min}$.

In the middle and right panels of Fig.~\ref{fig:VarMeff:PnMnMR} we show the NS mass versus the central density and the NS radius, respectively. The dotted segments between vertical dashes  correspond to the central densities $n_{\rm cen}\in [n_1^{\rm (MC)},n_2^{\rm (MC)}]$ within the Maxwell construction region. There are no stable star configurations with such central densities.
The maximum NS masses for $\Phi_{\rho,\rm min} = 0.3$, 0.5, 0.7 are 2.06, 2.16, and $2.21\, M_\odot$, respectively, proven to be above the  value $2.03M_\odot$  obtained with the original $\rho$ mass scaling $\Phi_\rho(f)=1-f$. With an increase of $\Phi_{\rho,\rm min}$ the value $n_{c,\rho}^{(\rm I)}$  increases and its influence on the EoS becomes less pronounced. For $\Phi_{\rho,\rm min} = 0.7 $ the $\rho^-$ condensate does not appear in a NS in the framework of the model under consideration. The radius of the NS with maximum mass increases with increase of $\Phi_{\rho,\rm min}$.
Thus, we have shown that the dependence of the $\rho$ meson effective mass on the scalar field plays a crucial role.

Modification of the  dependence of the $\rho$ meson effective mass on the scalar field in the absence of the $\rho^-$ condensation does not change the EoS at any composition of BEM, because the meson effective mass enters the energy density (\ref{edensity}) only in a combination $\eta_\rho(f)$. However, the $\rho^-$ condensate quantities depend not only on $\eta_\rho(f)$ but  also on  $m_\rho^* (f)$.
For the given $\eta_\rho(f)$ the scaling function $\chi_{\rho N}(f)$  changes, if we alter the effective mass scaling function $\Phi_\rho(f)$. Using the function $f(n)$ for the given matter composition, we obtain the $\chi_{\rho N}$ as a function  of the density.
\begin{figure}
	\includegraphics[width=\textwidth]{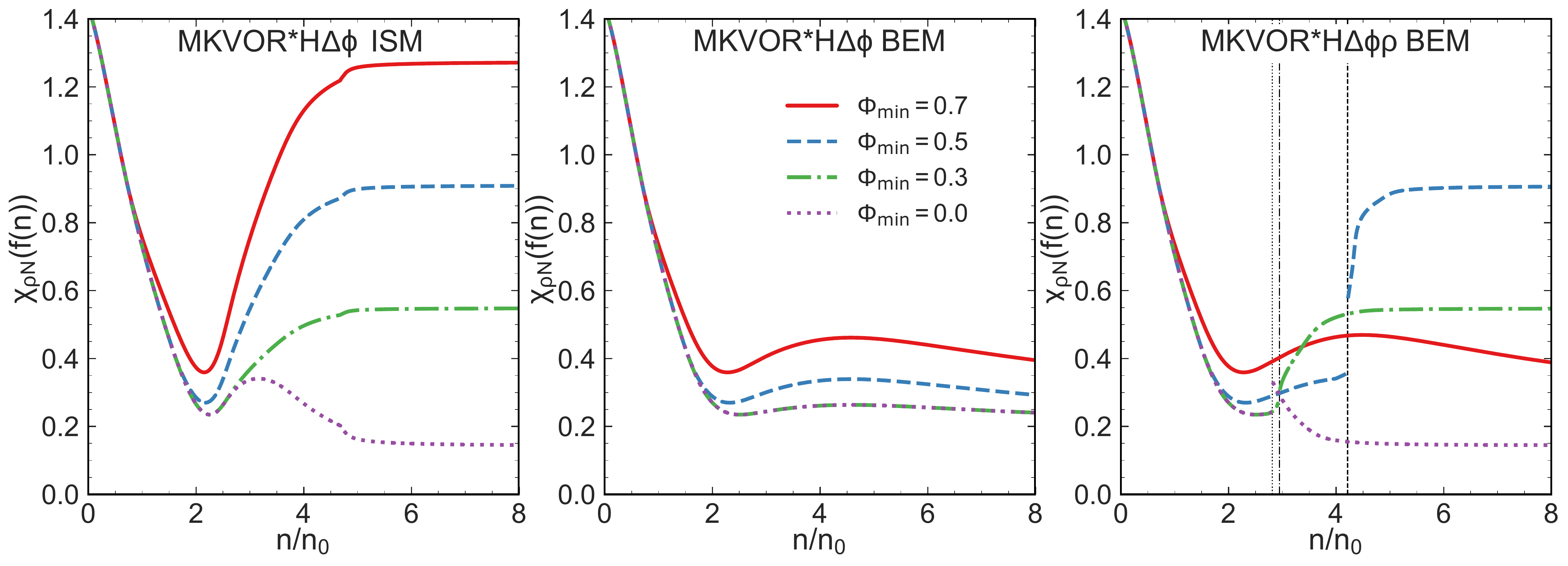}  
	\caption{
Scaling functions $\chi_{\rho N}(f)$ for MKVOR*H$\Delta\phi$ model in the ISM (left panel) and in the BEM (middle panel) and for MKVOR*H$\Delta\phi\rho$ model in BEM (right panel) for $\Phi_\rho(f)$ given by Eq.~(\ref{app:Phir}) with $\Phi_{\rho,\rm min} = 0.3$, 0.5, and 0.7. Lines labeled $\Phi_{\rho,\rm min}=0$ correspond to the scaling $\Phi_\rho = 1-f$. Vertical lines on the right panel denote the densities $n_{\rm c,\rho}^{(\rm I)}$. }
	\label{fig:VarMeff_chir}
\end{figure}

In Fig.~\ref{fig:VarMeff_chir} we show the resulting scaling functions $\chi_{\rho N}(f(n))$ as  functions of the density  for the
MKVOR*$\Delta$ model in ISM (left panel), for the MKVOR*H$\Delta\phi$ model in BEM (middle panel) and for the MKVOR*H$\Delta\phi\rho$ model in BEM (right panel), for parameters $\Phi_{\rho,\rm min}=  0.3$, 0.5, and 0.7 and for the original scaling $\Phi_\rho = 1-f$. The sharp change of $\chi_{\rho N}(f(n))$ for the MKVOR*$\Delta$ model in ISM at $n \simeq 4.7 \, n_0$ is a consequence of the appearance of $\Delta$s in ISM, cf. \cite{MKVOR-Delta}. We see that for ISM in MKVOR*$\Delta$ model  saturation of the $\rho$ meson effective mass decrease results in an increase of $\chi_{\rho N}(f(n))$. In MKVOR*H$\Delta\phi$ model it also leads to the increase of $\chi_{\rho N}(f(n))$ in a density interval.  The lines for $\Phi_{\rho,{\rm min}}=0$ and $\Phi_{\rho,{\rm min}} = 0.3$ coincide, since without $\rho^-$ condensation the value $f(n)$ does not exceed $0.7$ in this model. With the $\rho^-$ condensation included in the MKVOR*H$\Delta\phi\rho$ model (right panel), the line for $\Phi_{\rho,{\rm min}}=0.7$ does not change (the condensate is absent), and $\chi_{\rho N}(f(n))$ curves for $\Phi_{\rho,{\rm min}} = 0, 0.3, 0.5$ acquire jumps to another branch. On the new branch at $n = 8 \, n_0$ they reach values $\chi_{\rho N}(f(n = 8\, n_0)) = 0.15, \, 0.55, \, 0.91$ for $\Phi_{\rho,{\rm min}} = 0, 0.3, 0.5$, respectively.

\subsection{Variation of $\eta_\rho(f)$}\label{sec:var-etar}

In Section \ref{sect:MKVOR} we have shown that for the MKVOR*H$\Delta\phi\rho$ model the maximum NS mass  is practically insensitive to the choice of the tail of $\eta_\rho(f)$ function. Here we investigate how a stronger variation of $\eta_\rho(f)$ for $f$ larger than some value $\tilde{f}_{\rho}$  affects the EoS with $\Delta$s and  $\rho^-$ condensate. To preserve the same EoS as we have used  at  densities $n\lsim 2\, n_0$ (for $f<\tilde{f}_{\rho}$) we choose the value $\tilde{f}_\rho =0.45$.
\begin{figure}
	\centering
	\includegraphics[width=.45\textwidth]{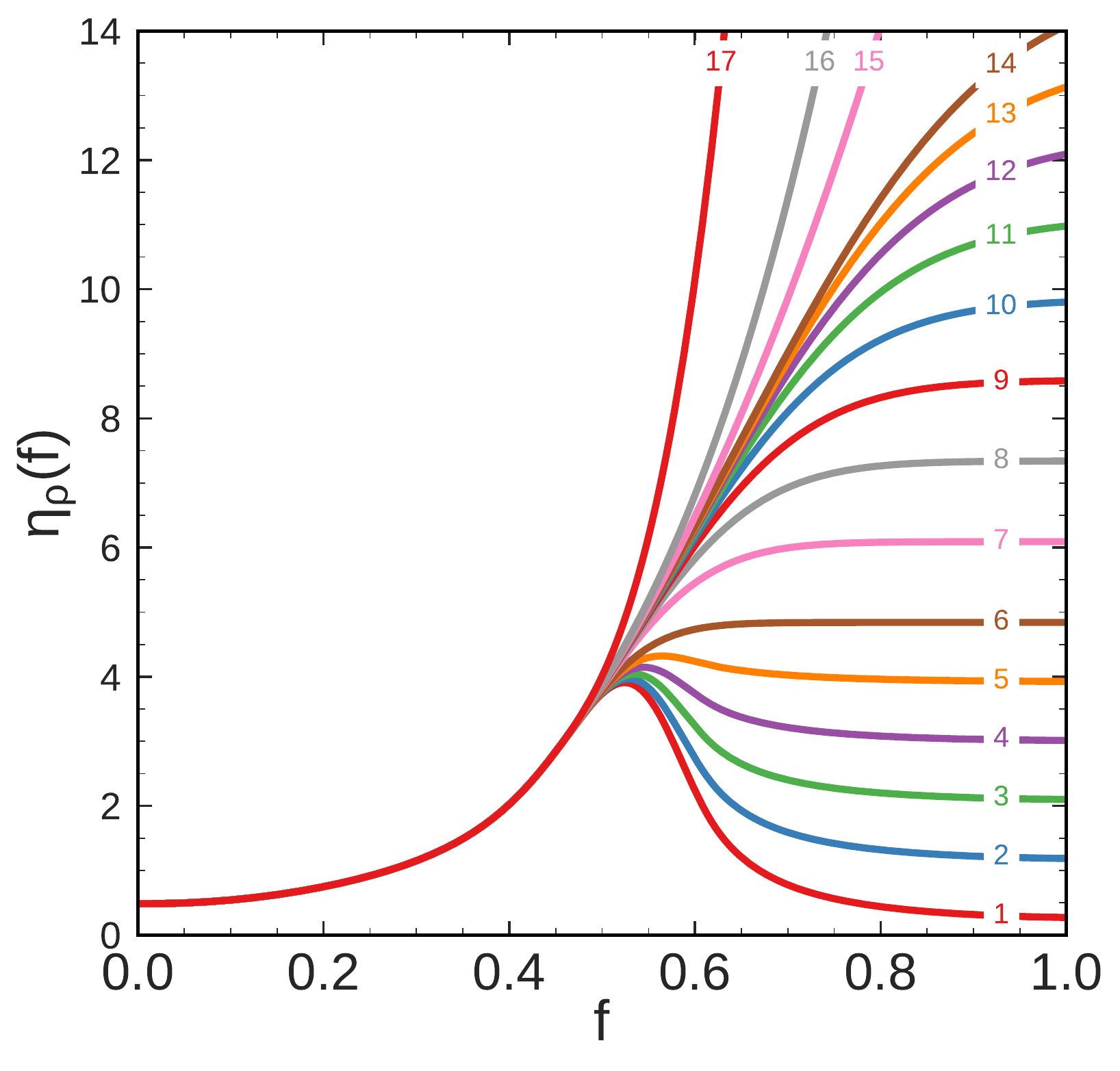} 
	\includegraphics[width=.45\textwidth]{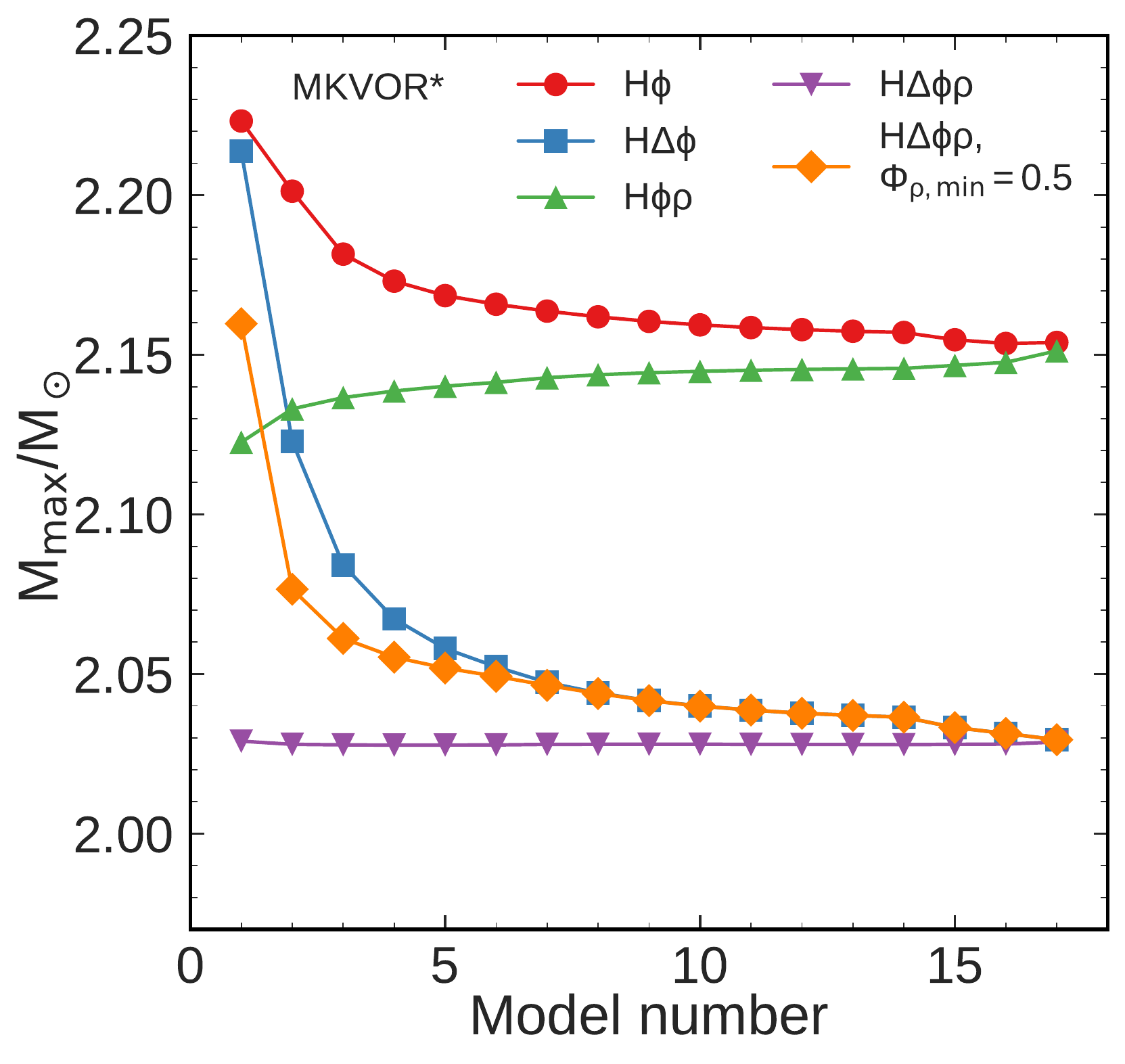} 
\caption{Left panel: 17 variations  of the scaling functions $\eta_\rho(f)$, given by Eqs. (\ref{eq:etar_a}) and (\ref{eq:eta_r_j}) labeled from 1 to 17.
Right panel: maximum NS mass versus a number of the model (from 1 to 17) of $\eta_\rho(f)$ for MKVOR*[H$\phi$, H$\Delta\phi$, $H\phi\rho$, H$\Delta\phi\rho$] models.}
	\label{fig:VarEta_eta}
\end{figure}

The family of various $\eta_\rho(f)$ functions, numerated from 1 to 17, which we would like to explore in this section is presented on the left panel in Fig.~\ref{fig:VarEta_eta}. On the right panel  we show the  maximum NS masses for various models plotted versus the label of the model from 1 to 17.

First consider models exploiting universal scaling of hadron masses.
We see that for the MKVOR*H$\phi$ (curve with circles on right panel) and MKVOR*H$\Delta\phi$ (curve with squares) models with the  $\rho^-$ condensation switched off the maximum NS mass monotonously decreases when we change the $\eta_\rho(f)$ from model  1 to 17. The largest values of $M_{\rm max}$ are realized for model 1. In this case the stiffening mechanism proposed in~\cite{Maslov:cut}, which damps a growth of the scalar field with increasing density, is the most efficient. This circumstance justifies our choice of the $\eta_\rho(f)$ for the MKVOR*-based models, as it maximizes the NS mass for models without inclusion of the $\rho^-$ condensate.

If a possibility of the $\rho^-$ condensate  is included but $\Delta$s are not incorporated, see the curve for the MKVOR*H$\phi\rho$ model (triangles),  the trend is opposite: the maximum NS mass increases with an increase of the model number (from 1 to 17).
This occurs because with the increase of $\eta_\rho (f)$ the amplitude of $\rho^-$ condensate decreases, thereby reducing the softening of the EoS.

A peculiar situation occurs, if the $\rho^-$ condensate is included in the model with $\Delta$ baryons, see the curve  in Fig.~\ref{fig:VarEta_eta} for the MKVOR*H$\Delta\phi\rho$ model labeled by returned-triangles.
The maximum NS mass proves to be almost independent on the choice of the model (from 1 to 17) for the $\eta_\rho$ function. The increase of $\eta_\rho$ function (increase of the model number) reduces  the $\rho^-$ condensate, but its softening effect on the EoS is taken over by the $\Delta$ baryons, which population increases with the model number. This compensation effect is incidental and can be removed, e.g., by a variation of  the $\rho$ meson mass scaling function. Diamond-labeled curve on the right panel in Fig.~\ref{fig:VarEta_eta} shows $M_{\rm max}$ for various $\eta_\rho$ models (from 1 to 17) for the model MKVOR*H$\Delta\phi\rho$ with the effective $\rho$ mass scaling function given by Eq.~(\ref{app:Phir}) with $\Phi_{\rho,\rm min}=0.5$ (dash curve  on left panel of  Fig.~\ref{fig:meff_conc}). We see that the pattern is now similar to that for the MKVORH$\Delta\phi$ model, since the $\rho^-$ condensation effect is smaller for models labeled 1 to 6 and vanishes completely for models from 7 to 17.

\section{Conclusion}

The charged $\rho^-$ meson condensation in dense isospin-asymmetric matter was proposed in~\cite{v97}. It may occur provided $\rho$ mesons are treated as non-Abelian  bosons and effective $\rho$ meson mass decreases with increase of the baryon density. This phenomenon was then studied in~\cite{Kolomeitsev:2004ff} within the KVOR model in the beta-equilibrium  matter. In this paper we studied the phenomenon of the charged $\rho$ condensation in the framework of the KVORcut03- and MKVOR*-based models developed in~\cite{Maslov:2015msa,Maslov:2015wba,MKVOR-Delta}, where we exploit a new mechanism~\cite{Maslov:cut} for stiffening of the equation of state (named the cut-mechanism). They include hyperons and $\Delta$ baryons and nevertheless satisfy known experimental constraints put on the equation of state from various analyses of atomic nuclei, heavy-ion collisions and  neutron stars.

We demonstrated that in the KVORcut03 model exploiting the cut mechanism in $\omega$ sector the $\rho^-$ condensation  appears  in the beta equilibrium matter by the second-order phase transition for densities $n>4.6\,n_0$, see Fig.~\ref{fig:cut03_nr}. Influence of the $\rho^-$ condensate on the equation of state is moderate and the maximum neutron star mass reduces only slightly, from 2.17\,$M_\odot$ to 2.16\,$M_\odot$. In the presence of hyperons, within the KVORcut03H$\phi$, KVORcut03H$\phi\sigma$, KVORcut03H$\Delta\phi$ and KVORcut03H$\Delta\phi\sigma$ models, the $\rho^-$ condensate does not appear in the beta equilibrium matter.

In the MKVOR*H$\Delta\phi$ model~\cite{MKVOR-Delta}, which is a version of the MKVOR model proposed in~\cite{Maslov:2015msa,Maslov:2015wba} extended to appropriately include $\Delta$ isobars (with the potential $U_\Delta (n=n_0)=-50$\,MeV in the isospin symmetrical matter), the cut-mechanism is embedded in the $\rho$ and $\om$ sectors. The model turns to be  sensitive to inclusion of the $\rho^-$ condensation. The $\rho^-$ condensate appears in this (MKVOR*H$\Delta\phi\rho$) model at density $n_{\rm c\rho}^{\rm (II)}=2.74\, n_0$ in a second-order phase transition. However already at smaller densities there exists a second solution of the equations for the scalar field $f(n)$ and particle fractions with a very different particle composition. This new solution is energetically unfavorable at the density of its appearance in comparison with the first solution, which develops continuously from a small condensate fraction. Only at a density $n_{\rm c,\rho}^{\rm (I)}$ the second solution becomes energetically favorable, and the system, if considered at fixed density, would undergo a first-order phase transition to a state with a large fraction of the $\rho^-$ condensate. The  density $n_{\rm c,\rho}^{\rm (I)}$ and the resulting particle concentrations are sensitive to variation of the $\eta_\rho(f)$ scaling function, so $n_{\rm c,\rho}^{\rm (I)}$ varies between 2.76 and 2.83\,$n_0$ for various tails of the $\eta_\rho$ function that have been exploited in~\cite{MKVOR-Delta}.
The appearance of the $\rho^-$ condensate leads to  occurrence  of $\Lambda$s and $\Delta^-$s already in neutron stars with  masses above $M\simeq 1.14 M_{\odot}$ and the new $\Delta^0$ and $\Delta^+$ species appear in heavier neutron stars (see Fig.~\ref{fig:MKVtails-conc} and Table~\ref{tab:MKVOR-crit-param}). The direct Urca reactions on $\Delta^-$ may occur only at densities exceeding the critical value for the process on nucleons but direct Urca reactions on $\Lambda$s prove to be allowed in neutron stars with masses as low as $1.14 M_{\odot}$. A feasible problem with description of the neutron star cooling might be avoided provided the $\Lambda$ Cooper pairing gaps are sufficiently high.

The appearance of the $\rho^-$ condensate leads to a strong reduction of the pressure and  softening of the EoS, see Fig.~\ref{fig:MKVtails-press}. Although the particle composition of the beta equilibrium matter depends strongly on the tail of the $\eta_\rho$ function, which we exploit, see Fig.~\ref{Fig-1-new}, the resulting equation of state exhibits only a small difference at densities $n \gsim 3.4 \, n_0$. The particle concentrations and the pressure for $2.37 \, n_0 \lsim n \lsim 3.4 \, n_0$ are different for various tails, but for describing the neutron star structure the pressure in this region is to be replaced by the Maxwell construction. On the Maxwell constructions the pressure proves to be weakly dependent on the tail. Consequently, the maximum mass of the neutron star is 2.03\,$M_\odot$ independently of the tail, which is substantially smaller than that for the MKVOR*H$\Delta\phi$ model, $M_{\rm max}=2.21\,M_\odot$, but still remaining within the error-bars of the current most precise measurements of heaviest neutron stars~\cite{Demorest:2010bx,Antoniadis:2013pzd}.

A possibility of the $\rho^-$ condensation at densities reachable in neutron stars hinges on reduction of the $\rho$ meson mass in medium. We varied the scaling function $\Phi_\rho=m_\rho^*/m_\rho$ for the effective $\rho$ meson mass, preventing the $\Phi_\rho$ to become smaller than a value $\Phi_{\rho,\rm min}$, see Fig.~\ref{fig:meff_conc}. It turns out that if $\Phi_{\rho,\rm min}\gsim 0.5$ then the influence of the $\rho^-$ condensation proves to be minor, and the maximum NS mass differs from that for the MKVOR*H$\Delta\phi$ model maximally by  2\%, see Fig. \ref{fig:VarMeff:PnMnMR}. For $\Phi_{\rho,\rm min}\gsim 0.7$ the critical density of the $\rho^-$  condensation exceeds the maximum density possible in neutron stars.
So the variation of $\Phi_\rho$ provides an effective mechanism to control  the presence and the magnitude of the $\rho^-$ condensate.

 Also, we investigated various choices of the $\eta_\rho$ function and its influence on the maximum mass of a neutron star. We demonstrated that the choice of $\eta_\rho$ proposed in~\cite{Maslov:2015msa,Maslov:2015wba} for the models without $\rho^-$ condensate is the optimal one leading to  increase of the maximum neutron star mass, see Fig.~\ref{fig:VarEta_eta}. In this case $\eta_\rho(f)$ is an increasing function of $f$ for $f \lsim 0.5$  (we use it in order to decrease the proton concentration in BEM increasing, thereby, the direct Urca reaction threshold), and then $\eta_\rho(f)$ decreases sharply for $f \gsim 0.5$  that allows to stiffen  the EoS owing to the cut-mechanism described in ~\cite{Maslov:cut}.
In the presence of the $\rho^-$ condensate an peculiar situation occurs: for the particular choice of the $\rho$ meson mass scaling and the $\Delta$ potential in isospin symmetric matter the maximum neutron star mass turns out to be practically independent of the various choices of the $\eta_\rho$ function we have studied. Although the $\rho^-$ condensation impact on the EoS changes while we vary $\eta_\rho$, this change proves to be compensated by the corresponding change in the $\Delta^-$ concentration.

Concluding, the influence of the $\rho^-$ condensation on the equation of state proves to be strongly model dependent. Nevertheless in the models we have studied including the hyperon and $\Delta$ isobar degrees of freedom as well as the charged $\rho$ condensation the observational constraint on the maximum neutron star mass is  fulfilled.

\section*{Acknowledgements}

We thank M.~Borisov and F.~Smirnov for the interest in this work and help.
The reported study was funded by the Russian Foundation for Basic Research (RFBR) according to the research project No 16-02-00023-A. The work was also supported by the Ministry of Education and Science of the Russian Federation within the state assignment,
project No~3.6062.2017/BY, by the Slovak grant No.~VEGA-1/0469/15, by ``NewCompStar'', COST Action MP1304, and partially (Sect. 4) by the
Russian  Science  Foundation,  grant  No.~17-12-01427. K.A.M. acknowledges the support by grant of the Foundation for the Advancement of Theoretical Physics ``BASIS".
E.E.K acknowledges the support by grant of the Plenipotentiary of the Slovak Government to JINR.

\appendix {\bf Appendix}

\section{Lagrangian of the model}\label{app:Lag}

Here we present the  Lagrangian of  our model, which includes baryons
$b = (n,p,\Lambda,\Sigma^{\pm,0},\Xi^{-,0};$ $\Delta^{\pm,0,++})$ coupled to meson mean fields, $m=(\sigma, \omega, \rho,\phi)$,  $\sigma$ is the scalar meson  and $\omega, \rho$, and $\phi$ are vector mesons, and leptons $l$,
\begin{align}\label{lagr}
\mathcal{L}&=\sum_{b}\mathcal{L}_{b}+\mathcal{L}_{\sigma}+\mathcal{L}_{\om}+\mathcal{L}_{\rho}
+\mathcal{L}_{l}\,,
\end{align}
where
\begin{align}
\mathcal{L}_{b}&= \overline{\Psi}_{b}\big[i\gamma^\mu D_\mu - m_b\Phi_b\big] \Psi_b\,,
\nonumber\\
D_\mu &= \partial_\mu + i g_{\om b} \chi_{\om b} \omega_\mu
+  i g_{\rho b} \chi_{\rho b}\vec{t}_b \vec{\rho}_\mu + i g_{\phi b} \chi_{\phi b} \phi_\mu
\nonumber
\end{align}
for both spin-1/2 and spin-3/2 baryons taking  into account different spin degeneracy factors, 2 and 4 respectively. The Lagrangian for $\sigma$, $\om$, $\rho$, and $\phi$ mesons is
\begin{align}
\mathcal{L}_{\sigma}  &= \frac12 \partial_\mu \sigma \partial^\mu \sigma
- \frac12 m_\sigma^{2}\Phi_\sigma^2 \sigma^2  - {U}(\sigma)\,,
\nonumber\\
\mathcal{L}_{\om}  &=
-\frac14(\partial_\mu \om_\nu - \partial_\nu \om_\mu)^2 + \frac12 m_\omega^{2}\Phi_\omega^2\, \om_\mu \om^\mu\,,
\nonumber\\
\mathcal{L}_{\phi}  &=
- \frac14 (\partial_\mu \phi_\nu - \partial_\nu \phi_\mu)^2 + \frac12 m_\phi^{2}\Phi_\phi^2\, \phi_\mu \phi^\mu\,,
\nonumber\\
\mathcal{L}_{\rho}  &=
 - \frac14\big(\partial_\mu \vec{\rho}_\nu -  \partial_\nu \vec{\rho}_\mu + g_\rho' \chi_\rho' [\vec \rho_\mu \times \vec \rho_\nu]\big)^2
+ \frac12 m_\rho^{2}\Phi_\rho^2 \vec{\rho}_\mu \vec{\rho}^{\,\mu}\,.
\label{Lag-mes}
\end{align}
We introduced the mass and coupling scaling functions ($\Phi$ and $\chi$, $\chi^{'}$) in the Lagrangian. $U(\sigma)=bf^3/3+cf^4/4$ is the $\sigma$ field dependent potential, $f=g_{\sigma N}\chi_{\sigma N}\sigma/m_N$. The $\rho$ meson term is written following the hidden local symmetry principles~\cite{Bando,Bando1}, which imply $g_{\rho N}=g'_\rho=g_\rho$.

One further puts $\om_\mu =\delta_{\mu 0}\om_0$ and $\phi_\mu =\delta_{\mu 0}\phi_0$, since it is easy to show that vector components
of these fields lead to an increase of the energy.

\section{ Parameters of the models and scaling functions}\label{app:scale-fun}

The KVORcut03 model, see \cite{Maslov:2015wba}, is an extension of the previously considered KVOR model with a  sharply varying function introduced in the $\om$ channel. The scaling functions entering the energy functional~(\ref{edensity}) are as follows:
\begin{eqnarray}
&&\eta^{\rm KVORcut}_\sigma(f) =  1 + 2 \frac{C_\sigma^2}{ f^{2}}\,  \big(\frac{b}{3} f^3 + \frac{c}{4} f^4\big) \,,\quad
\nonumber\\
&&\eta^{\rm KVORcut}_\omega(f) =  \Big[\frac{1 + z \bar{f}_0}{1 + z f}\Big]^\alpha + a_\omega \theta_{b_\om}(f-f_\om)
\,,
\label{eta-KVORcut}\\
&&\eta^{\rm KVORcut}_\rho (f) =
\Big[1 + 4\,\frac{C_\om^2}{C_\rho^2}\,\frac{(\bar{f}_0-f)\,z}{1+\bar{f}_0\,z}\Big]^{-1}\,.\quad
\nonumber
\end{eqnarray}
where $\bar{f}_0=f(n_0)=1-m_N^*(n_0)/m_N$.
We introduced here the switch functions
\begin{eqnarray}
\theta_y(x)=\frac12 \big[1+ \tanh(y x)\big]\,.
\label{eq:theta}
\end{eqnarray}
Parameters of the KVORcut03 model are collected in Table~\ref{tab:param-KVORcut03}.

\begin{table*}
	\caption{Parameters of the  KVORcut03 model}
	\begin{center}
		\begin{tabular}{lccc ccc ccc }
			\hline\hline
$C_\sigma^2$ & $C_\om^2$ & $C_\rho^2$ & $b \cdot 10^3$ & $c \cdot 10^3$  & $\alpha$ & $z$& $a_\om$  &$b_\om$ & $f_\om$ \\   
\hline
179.56& 87.600& 100.64& 7.7354& 0.34161& 1 & 0.65& $-$0.5 & 46.78 & 0.365 \\
			\hline \hline
		\end{tabular}
	\end{center}\label{tab:param-KVORcut03}
\end{table*}

The model MKVOR was proposed in~\cite{Maslov:2015msa,Maslov:2015wba}. It contains a sharply varying function  in the $\rho$-meson sector. In~\cite{MKVOR-Delta} it  was slightly modified by changing the tail of the $\eta_\rho$ function to get rid off the multiple solutions of the equation of motion for the $f(n)$ function in BEM and by changing the $\eta_\om$ function to avoid vanishing of the effective nucleon mass in ISM in  presence of $\Delta$ baryons. The resulting model, named MKVOR*, has the following scaling functions:
\begin{align}
\eta^{\rm MKVOR^*}_\sigma(f) &= \Big[1 - \frac{2}{3} C_\sigma^2 b f -
\frac{1}{2} C_\sigma^2 \Big(c -
\frac{8}{9} C_\sigma^2 b^2\Big) f^2 + \frac{1}{3} d f^3\Big]^{-1} \,,
\nonumber\\
\eta^{\rm MKVOR^*}_\omega(f)  &=\eta^{\rm KVORcut}_\omega(f)\,
\theta_{b_\om^*}(f_\om^*-f)
+ \frac{c_\om^*}{(f/f_\om^*)^{\alpha^*_\om}+1} \, \theta_{b_\om^*}(f-f_\om^*)\,,
\label{eta-MKVOR*}\\
\eta^{\rm MKVOR^*}_\rho(f) &=
\left\{
\begin{array}{lc}
\tilde{\eta}_\rho(f) \,, & f\le f_\rho^*\\
{}[c_0 + c_1 z + c_2 z^2 + c_3 z^3 + c_4 z^4]^{-1} \,, & f >f_\rho^*
\end{array}
\right.\,,
\quad  z=f/f_\rho^*-1\,, 
\nonumber
\end{align}
where $\widetilde{\eta}_\rho(f)$ is the scaling function introduced in~\cite{Maslov:2015wba}
\begin{align}
\tilde{\eta}_\rho(f) &= a_\rho^{(0)} + a_\rho^{(1)} f +
\frac{a_\rho^{(2)} f^2}{1 + a_\rho^{(3)} f^2}  +
\beta \exp\big(- \Gamma(f)(f - f_\rho)^2 \big),
\label{etarho-orig}\\
&\qquad \Gamma(f) = \gamma
\Big[1 + \frac{d_\rho (f-\bar{f}_0)}{1 + e_\rho (f-\bar{f}_0)^2} \Big]^{-1}\,,
\nonumber
\end{align}
and is extended by various tails parameterized in terms of coefficients $c_i$, three of which are fixed by the requirements of the continuity of the second derivative at the matching point $f=f^*_\rho$:
\begin{gather}
c_0 = \widetilde{\eta}^{-1}_\rho(f^*_\rho), \quad
c_1 = -f^*_\rho\,\widetilde{\eta}_\rho'(f^*_\rho)\,c_0^2,\quad
c_2 =  c_1^2/ c_0 - c_0^2 \widetilde{\eta}_\rho''(f^*_\rho)\, f^{*2}_\rho /2\,.
\nonumber
\end{gather}
Other parameters, $c_3$ and $c_4$, control the slope of the tail of the scaling function.
The original parametrization from~\cite{Maslov:2015wba} was labeled in~\cite{MKVOR-Delta} as ``tail 1'', and several other choices were proposed
\begin{align}
&\mbox{tail 2}: c_3=-10\,,\quad c_4=0\,;
\nonumber\\
&\mbox{tail 3}: c_3=0\,,\phantom{-1}\quad c_4=0\,;
\label{tail123}\\
&\mbox{tail 4}: c_3=0\,,\phantom{-1}\quad c_4=100\,.
\nonumber
\end{align}
 Parameters of the MKVOR* model  are collected in Table~\ref{tab:param-MKVOR}.

\begin{table*}
	\caption{Parameters of the MKVOR* model}
	\begin{center}
		\begin{tabular}{lccc ccc ccc}
			\hline\hline
$C_\sigma^2$ & $C_\om^2$ & $C_\rho^2$ & $b \cdot 10^3$ & $c \cdot 10^3$ &  $d$ &$\alpha$ & $z$& $a_\om$ & $b_\om$ \\
\hline
234.15&134.88&81.842&4.6750&$-$2.9742&$-$0.5&0.4&0.65& 0.11 &  7.1\\
\hline
$f_\om$	& $\beta$& $\gamma$ & $f_\rho$ & $a_\rho^{(0)}$& $a_\rho^{(1)}$ & $a_\rho^{(2)}$ & $a_\rho^{(3)}$ & $d_\rho$ & $e_\rho$ \\
			\hline
0.9 & 3.11 & 28.4 & 0.522 & 0.448 & $-$0.614 & 3   & 0.8 & $-$4 & 6 \\
\hline
$f_\rho^*$ &  $f_\om^*$ & $b_\om^*$ & $\alpha_\om^*$ & $c_\om^*$ & & & & &\\
\hline
0.62 & 0.95 & 100& 5.515 & 0.2299 & & & & &\\
			\hline \hline
		\end{tabular}
	\end{center}\label{tab:param-MKVOR}
\end{table*}

\section{Variations of  scaling functions $\Phi_\rho(f)$ and $\eta_\rho(f)$}

In Section ~\ref{sec:mrho},  we study  the dependence of the $\rho^-$ condensate on the $\rho$ meson scaling function $\Phi_\rho$. We use the following modification of the scaling function
\begin{align}
&\Phi_\rho(f)=\left\{
\begin{array}{ll}
1-f &,\, f\le f_s \\
(1-f_s)\,\big[1+\frac{\xi}{1+b_\rho\xi}\big(\frac{\xi}{2+b_\rho}-1\big)\big] &,\,f>f_s
\end{array}
\right.\,,\quad \xi=\frac{f-f_s}{1-f_s}\,.
\label{app:Phir}
\end{align}
The matching point  $f_s$ determined by the minimal value of the $\rho$ mass scaling function $\Phi_{\rho,\rm min}=\min_f(\Phi_\rho(f))$ is as follows
$$f_s=1-\Phi_{\rho,{\rm min}}-\delta\Phi_\rho,$$
with some off-set $\delta \Phi_\rho=0.1$ allowing for a smooth transition at $f=f_s$. The parameter $b_\rho=\frac{\Phi_{\rho,{\rm min}}}{\delta\Phi_{\rho}}-1$ is chosen so that $\Phi_\rho'(f=1)=0$.
 With $\Phi_{\rho,\rm min}=\delta\Phi=0$ we recover    $\Phi_\rho(f)=1-f$.

\begin{table}
	\centering
	\caption{Parameters of Eq. (\ref{eq:etar_a}), corresponding to lines 6--14 in Fig. \ref{fig:VarEta_eta}.}
\begin{tabular}{rrrrrrrrrr}
\hline\hline
\mbox{} & 6 & 7 & 8 & 9 & 10 & 11 & 12 & 13 &14\\\hline
$a_\rho$ & 2.00 & 3.25 &  4.50 & 5.75 & 7.00 & 8.25 & 9.50 & 10.75 & 12.00 \\\hline
$b_\rho$ & 4.2669 & 2.6258 & 1.8964 &	1.4841 & 1.2191 & 1.0344 & 0.8983 & 0.7938 & 0.7112  \\ \hline
$c_\rho$ & 3.4063 & 2.0896 & 1.5078 & 1.1795 & 0.9687 & 0.8218 & 0.7136 & 0.6306 & 0.5649 \\
\hline \hline
\end{tabular}
\label{tab:eta_r_a}
\end{table}

In Section~\ref{sec:var-etar} we investigate the influence of the choice of the scaling function $\eta_\rho(f)$ on the maximum neutron star mass.
In Fig.~\ref{fig:VarEta_eta} we consider 17 models of the $\eta_\rho(f)$ function, all of which coincide with original function (tail 2) given in Eq.~(\ref{eta-MKVOR*}) for $f<\tilde{f}_\rho$ and we fix $\tilde{f}_\rho =0.45$.
Models from 6 through 14 are given by the following expression
\begin{align}
\eta_\rho(f) =\left\{
\begin{array}{lcl}
\eta_{\rho, {\rm tail 2}}^{\rm MKVOR}(f) &, & f \le \tilde{f}_\rho \\
\eta_{\rho, {\rm tail 2}}^{\rm MKVOR} (\tilde{f}_\rho) + a_\rho \tanh(b_\rho \,\zeta + c_\rho \zeta^2) &,  & f >
\tilde{f}_\rho\end{array}
\right.\,,
\quad \zeta = \frac{f}{\tilde{f}_\rho} - 1\,.
\label{eq:etar_a}
\end{align}
The main parameter here is $a_\rho$, which controls the growth rate and value of $\eta_\rho(f)$ for $f \to 1$. Parameters  $b_\rho$ and $c_\rho$ depending on $a_\rho$ are chosen to obtain a smooth function with first and second continuous derivatives at $f=\tilde{f}_\rho$.
The parameters for models 6--14 shown in Fig.~\ref{fig:VarEta_eta} are listed in Table \ref{tab:eta_r_a}.

 Models 1--5 shown in the left panel of Fig.~\ref{fig:VarEta_eta} correspond to $j = 0$, 0.2, 0.4, 0.6, and 0.8. They are generated by the weighted averaging between the original $\eta_{\rho, {\rm tail 2}}^{\rm MKVOR}(f)$ and the lowest-lying $\eta_\rho(f)$ (model 6) given by (\ref{eq:etar_a}) for $a_\rho=2$:
\begin{align}
\eta^{(j)}_\rho(f) = (1-j) \eta_{\rho, {\rm tail 2}}^{\rm MKVOR}(f) + j \eta_\rho(f; a_\rho=2)\,.
\label{eq:eta_r_j}
\end{align}

Models 15, 16 and 17  shown on the left panel of Fig.~\ref{fig:VarEta_eta} correspond to $c_\rho^{(3)} = 10, 10^2, 10^3$, respectively. They are given by the parametrization

\begin{align}
\eta_\rho(f) =\left\{
\begin{array}{lcl}
\eta_{\rho, {\rm tail 2}}^{\rm MKVOR}(f) &, & f \le \tilde{f}_\rho \\
\eta_{\rho, {\rm tail 2}}^{\rm MKVOR} (\tilde{f}_\rho) + c_\rho^{(1)} \zeta + c_\rho^{(2)} \zeta^2 + c_\rho^{(3)} \zeta^3  &,  & f >
\tilde{f}_\rho\end{array}
\right.\,,\,\,\zeta = \frac{f}{\tilde{f}_\rho} - 1\,,
\label{eq:etar_c}
\end{align}
where coefficients $c_\rho^{(1)} = \tilde{f}_\rho \frac{\rmd}{\rmd f}\eta_{\rho, {\rm tail 2}}^{\rm MKVOR}(\tilde{f}_\rho)$, $c_\rho^{(2)} = \frac12 \tilde{f}_\rho^{2} \frac{\rmd^2}{\rmd f^2}\eta_{\rho, {\rm tail 2}}^{\rm MKVOR}(\tilde{f}_\rho)$ are chosen to preserve the smoothness of the function at $f=\tilde{f}_\rho$.


\begin{thebibliography}{99}
\def\shtitle#1{#1}

\bibitem{Lattimer:2012nd}
J.~M.~Lattimer,
Ann.\ Rev.\ Nucl.\ Part.\ Sci.\  {\bf 62} (2012) 485.

\bibitem{Woosley}
S. E.~Woosley, A.~Heger, and T. A.~Weaver,
Rev.\ Mod.\ Phys.\ {\bf 74} (2002) 1015.

\bibitem{Danielewicz:2002pu}
P.~Danielewicz, R.~Lacey, and W.~G.~Lynch,
Science {\bf 298} (2002) 1592.

\bibitem{Fuchs}
C.~Fuchs,
Prog.\ Part.\ Nucl.\ Phys.\ {\bf 56} (2006) 1.


\bibitem{Walecka1974}
J. D.~Walecka,
Ann. Phys. (N.Y.) {\bf 83} (1974) 491.

\bibitem{Boguta77}
J.~Boguta and A.R.~Bodmer,
Nucl. Phys. A {\bf 292} (1977) 413.
\bibitem{Boguta77a}J.~Boguta and  H.~St\"ocker,
 Phys. Lett. B {\bf 120} (1983) 289.
\bibitem{Boguta77b}
P.-G.~Reinhard, M.~Rufa, J.~Maruhn, W.~Greiner, and J.~Friedrich,
Z. Phys. A {\bf 323} (1986) 13.
\bibitem{Boguta77c}W.~Pannert, P.~Ring, and J.~Boguta,
Phys. Rev. Lett. {\bf 59} (1987) 2420.

\bibitem{SerotWalecka}
B. D.~Serot and J.D.~Walecka,
Adv. Nucl. Phys. {\bf 16} (1986) 1.
\bibitem{SerotWaleckaA} P.-G.~Reinhard,
Rep. Prog. Phys. {\bf 52} (1989) 439.

\bibitem{Glendenning} N. K.~Glendenning, {\em Compact Stars: Nuclear Physics,
Particle Physics, and General Relativity,} second ed.,
Springer-Verlag, New York, 2000.

\bibitem{Weber}  F. Weber, {\em Pulsars as
Astrophysical Laboratories for Nuclear and Particle Physics,} IoP
Publishing, Bristol, 1999.

\bibitem{Savushkin2015}
L. N. Savushkin,
Phys. Part. Nucl. {\bf 46} (2015) 859.


\bibitem{APR}
A.~Akmal, V. R.~Pandharipande, and D. G.~Ravenhall,
Phys.\ Rev.\ C {\bf 58} (1998) 1804.

\bibitem{FP}
B.~Friedman and V. R.~Pandharipande,
Nucl.\ Phys.\ A {\bf 361} (1981) 502.

\bibitem{Gandolfi:2009nq}
S.~Gandolfi, A.~Y.~Illarionov, S.~Fantoni, J.~C.~Miller, F.~Pederiva, and K.~E.~Schmidt,
Mon.\ Not.\ R.\ Astron.\ Soc.\ {\bf 404} (2010) L35.

\bibitem{Gandolfi12}
S.~Gandolfi, J.~Carlson, and S.~Reddy, Phys. Rev. C {\bf 85} (2012) 032081.
\bibitem{Gandolfi12a}
S.~Gandolfi, J.~Carlson, S.~Reddy, A.~W.~Steiner, and R.~B.~Wiringa,
Eur.\ Phys.\ J.\ A {\bf 50} (2014) 10.

\bibitem{Hebeler:2014ema}
K.~Hebeler and A.~Schwenk,
Eur.\ Phys.\ J.\ A {\bf 50} (2014) 11.

\bibitem{Tews13}
I.~Tews, T.~Kr\"uger, K.~Hebeler, and A.~Schwenk,
Phys.\ Rev.\ Lett.\  {\bf 110} (2013), 032504.


\bibitem{Lynch}
W. G.~Lynch, M. B.~Tsang, Y.~Zhang, P.~Danielewicz, M.~Famiano, Z.~Li, and A.W.~Steiner,
Prog. Part. Nucl. Phys. {\bf 62} (2009) 427.

\bibitem{Demorest:2010bx}
P.~Demorest, T.~Pennucci, S.~Ransom, M.~Roberts, and J.~Hessels,
Nature {\bf 467} (2010) 1081.

\bibitem{Fonseca}
  E.~Fonseca {\it et al.},
  Astrophys.\ J.\  {\bf 832}  (2016) 167.

\bibitem{Antoniadis:2013pzd}
J.~Antoniadis, P. C. C.~Freire, N.~Wex, T. M.~Tauris, R. S.~Lynch,
M. H.~van~Kerkwijk, M.~Kramer, and C.~Bassa,
Science {\bf 340} (2013) 6131.


\bibitem{Blaschke:2004vq}
D.~Blaschke, H.~Grigorian, and D. N.~Voskresensky,
Astron.\ Astrophys.\ {\bf 424} (2004) 979.

 \bibitem{Kolomeitsev:2004ff}
E. E.~Kolomeitsev and D. N.~Voskresensky,
Nucl.\ Phys.\ A {\bf 759} (2005) 373.

\bibitem{Grigorian:2016leu}
H.~Grigorian, D.~N.~Voskresensky, and D.~Blaschke,
Eur.\ Phys.\ J.\ A {\bf 52} (2016) 67.

\bibitem{Bogdanov:2012md}
S.~Bogdanov,
Astrophys.\ J.\  {\bf 762} (2013) 96.

\bibitem{Hambaryan2014}
V.~Hambaryan, R.~Neuh\"auser, V.~Suleimanov, and K.~Werner,
J. Phys.: Conf. Series {\bf 496} (2014) 012015.

\bibitem{Heinke:2014xaa}
C. O.~Heinke, H. N.~Cohn, P. M.~Lugger, N. A.~Webb, W. C. G.~Ho, J.~Anderson, S.~Campana,
S.~Bogdanov, D.~Haggard, A. M.~Cool, and J. E.~Grindlay,
Mon. Not. R. Astron. Soc. {\bf 444} (2014) 443.

\bibitem{Podsiadlowski}
P.~Podsiadlowski, J. D. M.~Dewi, P.~Lesaffre, J. C.~Miller, W. G.~Newton, and J. R.~Stone,
Mon. Not. R. Astron. Soc. {\bf 361} (2005) 1243.

\bibitem{Kitaura:2005bt}
F. S.~Kitaura, H. T.~Janka, and W.~Hillebrandt,
Astron.\ Astrophys.\  {\bf 450} (2006) 345.

\bibitem{Blaschke2016}D.~Alvarez-Castillo, A.~Ayriyan, S.~Benic, D.~Blaschke, H.~Grigorian,
and S.~Typel,
Eur.\ Phys.\ J.\ A {\bf 52}  (2016)  69.

\bibitem{Kolomeitsev2016} I.~Tews, J.~M.~Lattimer, A.~Ohnishi, and E.~E.~Kolomeitsev,
arXiv:1611.07133 [nucl-th].

\bibitem{Klahn:2006ir}
T.~Kl\"ahn, D.~Blaschke, S.~Typel, E. N. E.~van Dalen, A.~Faessler, C.~Fuchs, T.~Gaitanos, H.~Grigorian, A.~Ho, E. E.~Kolomeitsev, M. C.~Miller, G.~R\"opke, J.~Tr\"umper, D. N.~Voskresensky, F.~Weber, and H. H.~Wolter,
Phys.\ Rev.\ C {\bf 74} (2006) 035802.

\bibitem{Khvorostukhin:2006ih}
A. S.~Khvorostukhin, V. D.~Toneev, and D. N.~Voskresensky,
Nucl.\ Phys.\ A {\bf 791} (2007) 180.

\bibitem{Khvorostukhin:2008xn}
A. S.~Khvorostukhin, V. D.~Toneev, and D. N.~Voskresensky,
Nucl.\ Phys.\ A {\bf 813} (2008) 313.

\bibitem{Djapo:2008au}
J.~Schaffner-Bielich,
Nucl.\ Phys.\ A {\bf 804} (2008) 309;
H.~Djapo, B. J.~Schaefer, and J.~Wambach,
Phys.\ Rev.\ C {\bf 81} (2010) 035803.

\bibitem{Schulze-Rijken}
H.-J. Schulze and Th. Rijken,
Phys.\ Rev.\ C {\bf 84} (2011) 035801.


\bibitem{Drago2014}
A.~Drago, A.~Lavagno, G.~Pagliara, and D.~Pigato,
Phys.\ Rev.\ C {\bf 90} (2014) 065809.

\bibitem{Drago:2015cea}
A.~Drago, A.~Lavagno, G.~Pagliara, and D.~Pigato,
Eur.\ Phys.\ J.\ A {\bf 52} (2016) 40.


\bibitem{Long} W.~Long, J.~Meng, N.~Van~Giai, and S.-G.~Zhou,
Phys. Rev. C {\bf 69} (2004) 034319.


\bibitem{Typel}
S.~Typel and H. H.~Wolter,
Nucl. Phys. A {\bf 656} (1999) 331.

\bibitem{Hofmann}
F.~Hofmann, C. M.~Keil, and H.~Lenske,
Phys. Rev. C {\bf 64} (2001) 034314.

\bibitem{Niksirc}
T.~Nik\v{s}i\'c, D.~Vretenar, P.~Finelli, and P.~Ring,
Phys. Rev. C {\bf 66} (2002) 024306.

\bibitem{Lalazisis}
G.A.~Lalazissis, T.~Nik\v{s}i\'c, D.~Vretenar, and P.~Ring,
Phys. Rev. C {\bf 71} (2005) 024312.

\bibitem{Gaitanos}
T.~Gaitanos, M.~Di~Toro, S.~Typel, V.~Baran, C.~Fuchs, V.~Greco,
and H. H.~Wolter,
Nucl. Phys. A {\bf 732} (2004) 24.

\bibitem{RocaMaza:2011qe}
X.~Roca-Maza, X.~Vi\~nas, M.~Centelles, P.~Ring, and P.~Schuck,
Phys.\ Rev.\ C  {\bf 84} (2011) 054309.

\bibitem{Typel2005}
S.~Typel,
Phys.\ Rev.\ C {\bf 71} (2005) 064301.
\bibitem{Typel2005a}
M. D.~Voskresenskaya and S.~Typel,
Nucl.\ Phys.\ A {\bf 887} (2012) 42.


\bibitem{Metag} V. Metag, Prog. Part. Nucl. Phys. {\bf 30} (1993) 75;
V.~Metag,
Prog. Part. Nucl. Phys. {\bf 61} (2008) 245.

\bibitem{Rapp}
R.~Rapp and J.~Wambach,
Adv. Nucl. Phys. {\bf 25} (2000) 1.

\bibitem{Koch}
V.~Koch,
Int. J. Mod. Phys. E {\bf 6} (1997) 203.

\bibitem{BrownRho}
G. E.~Brown and M.~Rho,
Phys. Rev. Lett. {\bf 66} (1991) 2720.
\bibitem{BrownRhoA}
G. E.~Brown and M.~Rho,
Phys. Rep. {\bf 396} (2004) 1.

\bibitem{Maslov:2015msa}
K.~A.~Maslov, E.~E.~Kolomeitsev, and D.~N.~Voskresensky,
Phys.\ Lett.\ B {\bf 748} (2015) 369.

\bibitem{Maslov:2015wba}
K. A.~Maslov, E. E.~Kolomeitsev, and D.N.~Voskresensky,
Nucl.\ Phys.\ A {\bf 950} (2016) 64.

\bibitem{Maslov:cut}
K. A.~Maslov, E. E.~Kolomeitsev, and D.N.~Voskresensky,
Phys.\ Rev.\ C {\bf 92} (2015) 052801.


\bibitem{MKVOR-Delta} E. E. Kolomeitsev, K. A. Maslov, and D. N. Voskresensky,
Nucl.\ Phys.\ A {\bf 961} (2017) 106.

\bibitem{v97}
D. N.~Voskresensky,
Phys.\ Lett.\, B {\bf 392} (1997) 262.


\bibitem{Riek:2008uw}
  F.~Riek, M.~F.~M.~Lutz and C.~L.~Korpa,
  Phys.\ Rev.\ C {\bf 80} (2009) 024902.


\bibitem{Bando}
M. Bando, T. Kugo, S. Vehara, K. Yamawaki, and T. Yamagida, Phys. Rev. Lett. {\bf 54} (1983) 1215.
\bibitem{Bando1}
M. Bando, T. Kugo, and K. Yamawaki, Phys. Rep. {\bf 164} (1988) 217.

\bibitem{Migdal78qcd}
A. B. Migdal,
JETP Lett. {\bf 28} (1978) 35.

\bibitem{Linde79}
A. D. Linde,
Phys. Lett. {\bf 86} (1979) 39.

\bibitem{Chernodub}
M. N.~Chernodub,  Lect. Notes Phys. {\bf 871} (2013) 143.

\bibitem{Mallick:2014faa}
R.~Mallick, S.~Schramm, V.~Dexheimer, and A.~Bhattacharyya,
Mon.\ Not.\ Roy.\ Astron.\ Soc.\  {\bf 449} (2015)  1347.


\bibitem{Schulz:1983pz}
  H.~Schulz, D.~N.~Voskresensky and J.~Bondorf,
  Phys.\ Lett.\  {\bf 133B} (1983)  141.

\bibitem{Skokov:2009yu}
  V.~V.~Skokov and D.~N.~Voskresensky,
  Nucl.\ Phys.\ A {\bf 828} (2009) 401.

\bibitem{Skokov:2010dd}
  V.~V.~Skokov and D.~N.~Voskresensky,
  Nucl.\ Phys.\ A {\bf 847} (2010) 253.

  \bibitem{Voskresensky:1987ut}
D. N.~Voskresensky, A.V.~Senatorov, B.~K\"ampfer, and H.J.~Haubold,
Astrophys.\ Space Sci.\  {\bf 138} (1987) 421.

\bibitem{Haubold:1988uu}
H.~J.~Haubold, B.~K\"ampfer, A.~V.~Senatorov, and D.~N.~Voskresensky,
Astron.\ Astrophys.\  {\bf 191} (1988) L22.

\bibitem{Migdal:1990vm}
A. B. Migdal, E. E. Saperstein, M. A. Troitsky, and D. N. Voskresensky,
Phys.\ Rept. {\bf 192} (1990) 179.

\bibitem{Glend92} N. K. Glendenning, Phys. Rev. D {\bf 46} (1992) 1274.

\bibitem{Heiselberg} H. Heiselberg, C. J. Pethick, and E. F. Staubo, Phys. Rev. Lett. {\bf 70} (1993) 1355.

\bibitem{Voskresensky:2001jq}
  D.~N.~Voskresensky, M.~Yasuhira and T.~Tatsumi,
  Phys.\ Lett.\ B {\bf 541} (2002) 93.

\bibitem{Voskresensky:2002hu}
  D.~N.~Voskresensky, M.~Yasuhira and T.~Tatsumi,
  Nucl.\ Phys.\ A {\bf 723} (2003) 291.

\bibitem{Maruyama:2005tb}
  T.~Maruyama, T.~Tatsumi, D.~N.~Voskresensky, T.~Tanigawa, T.~Endo and S.~Chiba,
  Phys.\ Rev.\ C {\bf 73} (2006) 035802.


\bibitem{Prakash-DU}
M.~Prakash, M.~Prakash, J. M.~Lattimer, and C. J.~Pethick,
Astrophys. J. {\bf 390} (1992) L77.

\bibitem{Takatsuka:2005bp}
  T.~Takatsuka, S.~Nishizaki, Y.~Yamamoto, and R.~Tamagaki,
  Prog.\ Theor.\ Phys.\  {\bf 115} (2006) 355.

\bibitem{Voskresensky:1986af}
D. N.~Voskresensky and A. V.~Senatorov,
Sov.\ Phys.\ JETP {\bf 63} (1986) 885
[Zh.\ Eksp.\ Teor.\ Fiz.\  {\bf 90} (1986) 1505].

\bibitem{Straaten} S. van Straaten, E. C. Ford, M. van der Klis, M. M´endez, and P.
Kaaret, Astrophys. J. {\bf 540} (2000) 1049.

\bibitem{Ozel:2015fia}
F.~\"Ozel, D.~Psaltis, T.~Guver, G.~Baym, C.~Heinke, and S.~Guillot,
Astrophys.\ J.\  {\bf 820} (2016) 28.

\bibitem{Suleimanov:2016llr}
V. F.~Suleimanov, J.~Poutanen, J.~N\"attil\"a, J. J. E.~Kajava, M. G.~Revnivtsev, and K.~Werner,
Mon.\ Not.\ Roy.\ Astron.\ Soc.\  {\bf 466} (2017) 906.

\bibitem{Trumper}
J. E.~Tr\"umper, V.~Burwitz, F.~Haberl, and V. E.~Zavlin,
Nucl.\ Phys.\ B (Proc.\ Suppl.) {\bf 132} (2004) 560.


\bibitem{Lattimer:2013hma}
J. M.~Lattimer and A. W.~Steiner,
Astrophys.\ J.\  {\bf 784} (2014) 123.

\bibitem{Steiner:2015aea}
A. W.~Steiner, J. M.~Lattimer, and E. F.~Brown,
Eur.\ Phys.\ J.\ A {\bf 52} (2016) 18.

\end{thebibliography}
\end{document}